\documentclass{aa}

\usepackage{graphicx}
\pdfoutput=1
\usepackage{txfonts}
\usepackage{url}
\usepackage{natbib}
\usepackage{color}
\bibpunct{(}{)}{;}{a}{}{,}
\begin{document}
\title{Modelling the dust emission from dense interstellar clouds: disentangling the effects of radiative transfer and dust properties}

\author{N. Ysard\inst{\ref{inst1}}
\and M. Juvela\inst{\ref{inst1}}
\and K. Demyk\inst{\ref{inst2}}
\and V. Guillet\inst{\ref{inst3}}
\and A. Abergel\inst{\ref{inst3}}
\and J.-P. Bernard\inst{\ref{inst2}}
\and J. Malinen\inst{\ref{inst1}}
\and C. M\'eny\inst{\ref{inst2}}
\and L. Montier\inst{\ref{inst2}}
\and D. Paradis\inst{\ref{inst2}}
\and I. Ristorcelli\inst{\ref{inst2}}
\and L. Verstraete\inst{\ref{inst3}}
          }
\institute{Department of Physics, PO Box 64, FI-00014 University of Helsinki, Finland, \label{inst1}\email{nathalie.ysard@ias.u-psud.fr} 
\and IRAP, CNRS (UMR5277), Universit\'e Paul Sabatier, 9 avenue du Colonel Roche, BP 44346, F-31028 Toulouse cedex 4, France\label{inst2}
\and IAS, CNRS (UMR8617), Universit\'e Paris-Sud 11, B\^atiment 121, F-91400 Orsay, France\label{inst3}}

\abstract{Dust emission is increasingly used as a tracer of the mass in the interstellar medium. With the combination of Planck and Herschel observatories, we now have both the spectral coverage and the angular resolution required to observe dense and cold molecular clouds. However, as these clouds are optically thick at short wavelengths but optically thin at long wavelengths, it is tricky to conclude anything about dust properties without a proper treatment of the radiative transfer.}
{Our aim is to disentangle the effects of radiative transfer and dust properties on the variations in the dust emission at long wavelengths. This enables us to provide observers with tools to analyse the dust emission arising from dense clouds.}
{We model cylindrical clouds with visual extinctions between 1 and 20 magnitudes, illuminated by the standard interstellar radiation field, and carry out full radiative transfer calculations using a Monte Carlo code. Dust temperatures are solved using the DustEM code for amorphous carbons and silicates representative of dust at high Galactic latitude (DHGL), carbon and silicate grains coated with carbon mantles, and mixed aggregates of carbon and silicate. We also allow for variations in the optical properties of the grains with wavelength and temperature. We determine observed colour temperatures, $T_{colour}$, and emissivity spectral indices, $\beta_{colour}$, by fitting the dust emission with modified blackbodies using a standard $\chi^2$ fitting method, in order to compare our models with observational results.}
{Radiative transfer effects can explain neither the low $T_{colour}$, the increased submillimetre emissivity measured at the centre of dense clouds, nor the observed $\beta_{colour}-T_{colour}$ anti-correlation for the models considered. Adding realistic noise to the modelled data, we show that it is unlikely to be the only explanation of the $\beta_{colour}-T_{colour}$ anti-correlation observed in starless clouds, which may instead be explained by intrinsic variations in the grain optical properties with temperature. Similarly the higher submillimetre emissivity and the low $T_{colour}$ have to originate in variations in the grain optical properties, probably caused by their growth to form porous aggregates. We find it difficult to infer the nature of the grains from the spectral variations in their emission, owing to radiative transfer effects for $\lambda \lesssim 300 \; \mu$m, and to the mixture of different grain populations for longer wavelengths. Finally, the column density is underestimated when determined with modified blackbody fitting because of the discrepancy between $T_{colour}$ and the "true" dust temperature in the innermost layers of the clouds.}
{}

   \authorrunning{}
\titlerunning{Modelling the dust emission from dense interstellar clouds}

\maketitle
%

\section{Introduction}
\label{introduction}

Even if the main lines are understood, a number of questions remain regarding the processes that form a star from a cloud of gas and dust. Recent observations indicate that the Galactic interstellar medium (ISM) is organised as a network of elongated filaments \citep{MAMD2010, Andre2010, Juvela2010} and that pre-stellar cores seem to form preferentially inside the densest of these filaments \citep{Hill2011, Nguyen2011}. However, the details of either the formation of the filaments or their later fragmentation into dense molecular clumps are still poorly known.

The physical, chemical, and dynamical state of the filaments is expected to influence the star formation efficiency and the mass of the new-born stars. It is thus crucial to characterize the properties of the gas and in particular, in the context of the Planck and Herschel missions, of the dust grains that compose Galactic interstellar filaments. Indeed the combination of these two observatories offers both the spectral coverage and the angular resolution necessary to enable the detection and the characterisation of dense, molecular, and potentially pre-stellar clouds, through the emission of the cold dust they contain. Much can be learned about the cold clumps from this emission as the properties of the grains are known to vary strongly from the diffuse ISM to the centre of dense molecular clouds (Tab. 1). For instance, the temperature of the grains decreases from the diffuse to the dense medium \citep{Schnee2010}, while their submillimetre emissivity increases \citep{Cambresy2001, Stepnik2003, Ridderstad2006, Kiss2006, Lehtinen2007, Schnee2008, Juvela2011, PlanckAbergel2011}. There is also observational evidence that the small grains disappear towards dense molecular clouds, as the deficit of emission in the mid-IR \citep{Bernard1999, Stepnik2003, Flagey2009}. In addition, the ratio of total to selective extinction $R_V$ increases from a standard value of 3.1 in the diffuse medium to up to $\sim 5.5$ towards dense clouds \citep{Fitzpatrick1988, Mathis1989, Cardelli1991, Campeggio2007}, which can be explained by the disappearance of the small grains \citep{Kim1994}. This indicates that the grains grow from the diffuse medium to the centre of dense molecular clouds. It also appears to be increasingly clear, though still debated, that dust emission at long wavelengths is more complex than the usual modified blackbody at temperature $T$, with a constant emissivity spectral index $\beta \sim 2$ (Tab. 1). Observations towards the diffuse gas as well as the dense medium show an anti-correlation between $\beta$ and $T$ \citep{Dupac2003, Desert2008, Paradis2010, Veneziani2010, PlanckAbergel2011, PlanckMontier2011}. This behaviour can be explained by models of the physics of the emission of amorphous solids \citep{Meny2007, Paradis2011}, and is also corroborated by laboratory measurements on interstellar dust analogues of amorphous carbons \citep{Mennella1998} and amorphous silicates \citep{Agladze1996, Boudet2005, Coupeaud2011}. \citet{Malinen2011} also showed that a $\beta-T$ anti-correlation can be produced by a warm source inside the clouds. In addition, \citet{Shetty2009} found that this can naturally result from noise uncertainties and suggested that the anti-correlation may also be produced by temperature variations along the line-of-sight. Furthermore, laboratory measurements and models also predict spectral variations in the dust emissivity \citep{Boudet2005, Meny2007, Mennella1998, Coupeaud2011}, which have already been observed in the Galactic ISM \citep{Paradis2009, PlanckAbergel2011}.

Thanks to the new generation of instruments available, and in particular to the combination of Planck and Herschel data, we are now able to observe dense and cold molecular clouds, and to test detailed models of grain emission and growth. However, to describe these dense media, where the illuminated edge is much warmer than the shielded centre, detailed calculations of the radiative transfer are compulsory. The dust emission spectrum arising from dense regions has already been studied by various authors. For instance, \citet{Fischera2008} and \citet{Fischera2011} investigated the link between the dust emission spectrum in spherical, passively heated cores and their environment (radiation field, external pressure). They showed that the radiative transfer effects lead to an overestimate of the sizes of the cores and to an underestimate of their masses, in agreement with \citet{Evans2001, Stamatellos2003, Stamatellos2004} and \citet{Malinen2011}. These authors used various dust models and cloud geometries, either spheres or cylinders, and either symmetric or not, and considered them as embedded or directly submitted to the interstellar radiation field, but all concluded that the radiative transfer has a very strong impact on the observed dust emission. They showed that the shape of the emission profiles differs from the shape of the column density profiles, that the apparent temperature diverges from the dust physical temperature, and also that the sizes and the masses of the clouds cannot be accurately estimated by fitting the dust emission with a modified blackbody \citep{Stamatellos2010, Fischera2011, Malinen2011}. All these results emphasized the importance of a proper treatment of the radiative transfer if one wants to trace the variations in the dust properties from the diffuse to the dense medium, or otherwise the importance of using suitable corrections.

Therefore, the first question to answer concerns our ability to disentangle the effects of radiative transfer, i.e. the mixing of different dust temperatures along one line-of-sight, and the effects of the variations in the grain properties on the dust emission coming from dense clouds. Can we link the "true" dust temperature at the centre of the clouds to the average temperature we measure? How should the grains grow to have the low temperatures and the higher submillimetre emissivity they exhibit in dense clouds? Does the observed $\beta-T$ anti-correlation originate from intrinsic properties of the grains or is it an effect of only radiative transfer and/or noise in the data? What intrinsic relation would be needed to reproduce the observed $\beta-T$ anti-correlation, when the effects of radiative transfer are considered? What is the impact of the variations in the grain properties and of the radiative transfer on the quantities that are usually measured: column density and dust opacity? These are the main questions that we try to answer in this paper. Our aim is to give key information to the observers to be able to analyse quantitatively observations of dense starless clouds.

The paper is organised as follows. In Section \ref{models_methods}, we describe the dust emission model and the radiative transfer code we use throughout the paper. In Section \ref{section_diffuse}, we detail the characteristics of the dust emission from dense interstellar filaments when assuming that the grains have the same optical properties as in the high Galactic latitude diffuse ISM \citep{Compiegne2011}. We explore the effects of both radiative transfer and noise on the relation between the emissivity spectral index and the colour temperature. In Section \ref{section_grain_growth}, we present the effects of grain growth on the emerging dust emission. We consider two scenarios of growth, accretion and aggregation, and investigate the relation between the colour temperature and the grain physical temperature. In Section \ref{dust_intrinsic_properties}, we show the impact of intrinsic variations in the dust optical properties with wavelength and then with temperature. In Section \ref{column_density_estimates}, we test the accuracy of the estimate of the hydrogen column density by fitting the dust emission with a modified blackbody and investigate how the variations in the dust emissivity from diffuse to dense regions can be used to probe grain growth. Finally, we present in Section \ref{conclusion} our conclusions.

\begin{table*}
\label{table_bibliographie}
\centering
\caption{Characteristics of the clouds used to compare the models of dense molecular clouds described in the paper. The first column gives the name of the observed cloud, the second column the value of the visual extinction at the centre of the cloud, the third column the visual extinction around the cloud (extinction of the incident radiation field), the fourth column the lowest colour temperature measured in the cloud, the fifth column indicates whether the submillimetre emission of big grains is enhanced towards the cloud centre, the sixth column shows whether an anti-correlation between the emissivity spectral index and the colour temperature is observed, and the last column indicates the corresponding references.}
\begin{tabular}{lcccccl}
\hline
\hline
Object & Central $A_V$ & External $A_V$ & $T_{colour}$ (K) & Submm & $\beta-T$  & References \\
\hline
Filament in Taurus & $\leq 10$    & $\sim 0.4 - 0.5$ & $12.1^{+0.2}_{-0.1}$ & $\times 3.4^{+0.3}_{-0.7}$ & Yes & \citet{Stepnik2003, PlanckAbergel2011} \\
Filament in Musca  & $\sim 8$     & $\sim 0.3 - 0.4$ & $\sim 12.5$          & $\smallsetminus$           & Yes & \citet{Juvela2011} \\
Planck - G126.6+24.5 & $\sim 3.5$ & $\leqslant 1$    & $11.6 \pm 0.8$       & $\smallsetminus$           & Yes & \citet{PlanckRistorcelli2011} \\
Lynds 1780         & $4$          & $\sim 0.25$      & $14.9 \pm 0.4$ & $\sim \times 1.5$ & $\smallsetminus$   & \citet{Ridderstad2006, Ridderstad2010} \\
IC5146 (core \# 1) & $\sim 21$    & $\smallsetminus$ & $10.2 \pm 0.1$       & Yes & $\smallsetminus$ & \citet{Kramer2003} \\
\hline
\end{tabular}
\end{table*}

\section{Models and methods}
\label{models_methods}

\subsection{Dust emission model}
\label{dust_emission_model}

We use the dust model, DustEM\footnote{Available at \url{htpp://www.ias.u-psud.fr/DUSTEM}.}, described in \citet{Compiegne2011}. DustEM is a versatile numerical tool that computes dust emission and extinction as a function of the grain size distribution and optical properties. Our standard dust model is defined as the DustEM model for dust in the diffuse ISM at high Galactic latitude (DHGL according to the nomenclature of \citet{Compiegne2011}). Three dust populations are taken into account: interstellar polycyclic aromatic hydrocarbons (hereafter PAHs), amorphous carbons (small, hereafter SamC, and large, LamC), and the so-called astronomical silicates (hereafter aSil). The optical properties of the SamC and LamC grains come from the laboratory measurements of \citet{Colangeli1995}, who measured the opacity of a sample of $sp^2$-rich hydrogenated amorphous carbon for $\lambda \leqslant 1900$ $\mu$m \citep{Compiegne2011}. For longer wavelengths, the absorption efficiency is extrapolated using a single emissivity spectral index $\beta = 1.55$, equal to its average value for $800 \leqslant \lambda \leqslant 1900$ $\mu$m. The optical properties of the aSil population were chosen to reproduce the observations of the diffuse ISM \citep{Draine1984} and have a single emissivity spectral index equal to 2.11 in the far-IR and in the submillimetre\footnote{We do not use the update of the optical properties made by \citet{Draine2003}, who added a flattening of the emissivity spectral index in the submillimetre ($\beta = 2.11 \rightarrow 1.7$). Spectral variations in $\beta$ are addressed separately in Section \ref{spectral_variations}.}. Then, for the small grains (PAHs \& SamC), we use log-normal size distributions with central radii $a_0$ and widths $\sigma$. For the larger grains, we use a power-law distribution $a^{\alpha}$, starting at $a = 4$ nm, with an exponential cut-off $e^{-[(a-a_t)/a_c]^{\gamma}}$ for $a \geq a_t$ \citep{Weingartner2001, Compiegne2011}. The dust model abundances and the parameters of the size distributions are given in Tab. 2. We also consider the change in the grain optical properties towards dense molecular clouds, which are usually assumed to be caused by their growth by accretion and/or aggregation. The populations of evolved dust are described in Section \ref{section_grain_growth} and Tab. 2.

\begin{table}
\label{tableau_dustem}
\centering
\caption{Dust model abundances and size distribution parameters (see Section \ref{dust_emission_model} for details). $\rho$ is the grain mass density, $Y$ is the mass abundance per H, $\kappa_{250 \; \mu{\rm m}}$ is the opacity at 250 $\mu$m, and $\beta$ is the intrinsic opacity spectral index for $\lambda > 100 \; \mu$m for each dust population. For aggregates, the percentages give their degree of porosity (see Section \ref{coagulation} for details).}
\begin{tabular}{ccccccc}
\hline
\hline
\multicolumn{7}{c}{Small grains (DHGL)} \\
\hline
     & $\rho$     & $\sigma$ & $a_0$ & $Y$                   & $\kappa_{250 \; \mu{\rm m}}$ & $\beta$\\
     & (g/cm$^3$) &          & (nm)  & ($M_{dust}/M_H$)      & (cm$^2$/g)                   &        \\
PAH  & 2.24       & 0.40     & 0.64  & 7.80$\times 10^{-4}$  & 0.001                        &        \\
SamC & 1.81       & 0.35     & 2.00  & 1.65$\times 10^{-4}$  & 0.002                        & 1.55   \\
\hline
\hline
\multicolumn{7}{c}{Big grains (DHGL)} \\
\hline
     & $\rho$ & $\alpha$ & $a_c, a_t$ & $Y$                  & $\kappa_{250 \; \mu{\rm m}}$ & $\beta$ \\
     &        &          & (nm)       &                      &                              &         \\
LamC & 1.81   & -2.8     & 150.0      & 1.45$\times 10^{-3}$ & 0.014                        & 1.55    \\
aSil & 3.5    & -3.4     & 200.0      & 7.8$\times 10^{-3}$  & 0.034                        & 2.11    \\
\hline
\hline
\multicolumn{7}{c}{Accreted grains} \\
\hline
      & $\rho$ & $\alpha$ & $a_c, a_t$ & $Y$                 & $\kappa_{250 \; \mu{\rm m}}$ & $\beta$  \\
accC  & 1.81   & -2.8     & 150.0      & 1.6$\times 10^{-3}$ & 0.015                        & 1.55     \\
accSi & 3.22   & -3.4     & 200.0      & 8.6$\times 10^{-3}$ & 0.066                        & 1.52     \\
\hline
\hline
\multicolumn{7}{c}{Aggregates} \\
\hline
     & $\rho$ & $\alpha$ & $a_c, a_t$ & $Y$                  & $\kappa_{250 \; \mu{\rm m}}$ & $\beta$  \\
0\%  & 2.87   & -2.4     & 234.0      & 1.02$\times 10^{-2}$ & 0.111                        & 1.33     \\
10\% & 2.59   & -2.4     & 242.0      & 1.02$\times 10^{-2}$ & 0.140                        & 1.32     \\
25\% & 2.16   & -2.4     & 256.0      & 1.02$\times 10^{-2}$ & 0.208                        & 1.30     \\
40\% & 1.72   & -2.4     & 276.0      & 1.02$\times 10^{-2}$ & 0.331                        & 1.27     \\
\hline
\end{tabular}
\end{table}

\subsection{Radiative transfer model}
\label{description_CRT}

We use the Monte Carlo radiative transfer model described in \citet{Juvela2003} and \citet{Juvela2005}, CRT\footnote{Available at \url{http://wiki.helsinki.fi/display/~mjuvela@helsinki.fi/CRT}.} (Continuum Radiative Transfer). Regarding the cloud geometry, until now, two types of clouds could be used to perform the calculations: spheres, divided into discrete cells with constant density, which are concentric shells, or three-dimensional clouds, divided into cubic cells that have a constant density. The second case is time-consuming as no assumption is made about the symmetries. Observations of the Galactic ISM indicate that interstellar matter is mostly distributed along elongated filaments in both the dense and the diffuse medium \citep{Abergel2010, Andre2010, Juvela2010, Juvela2011, MAMD2010, Arzoumanian2011}. To account for this particular geometry, we implemented a third type of cloud in CRT: finite circular cylinders. Our cylinders are divided into layers perpendicular to the axis of symmetry and each layer is divided into concentric rings. Each concentric ring is thus a cell of the modelled cloud, except for the central cell of each layer, which is a disk. One discrete cell has a constant density but density can vary from one cell to the other in both directions, that is to say as a function of the height along the axis of symmetry and as a function of the radial position. 

The CRT model consists of two independent parts. First, the radiation field is estimated at each position inside the cloud using Monte Carlo methods as described in \citet{Juvela2003}. In the case of cylinders, we consider incoming and outcoming photons over the entire surface of the cloud, including the two caps. Second, using the radiation field and the selected dust model, dust temperature and emission are calculated. The CRT model allows us to add different populations of grains, with distinct spatial distributions. Dust temperature distributions are solved using DustEM \citep{Compiegne2011}. The non-isotropic and multiple scattering events are taken into account, as well as the re-emission by dust grains by iterating CRT \citep{Juvela2003, Juvela2005}.

\subsection{Definition of the test cloud}
\label{section_cloud_geometry}

In CRT, a cloud is defined by its hydrogen density distribution, $\rho(r)$. Cylindrical isothermal self-gravitating clouds have distributions varying as $\rho(r) \propto r^{-4}$ in the outer regions \citep{Ostriker1964}. However, interstellar filamentary clouds usually exhibit much shallower profiles from $\rho(r) \propto r^{-1.5}$ to $r^{-2.5}$, associated with a flat distribution at the centre \citep{Alves1998, Stepnik2003, Arzoumanian2011}. \citet{Fiege2000} demonstrated that these profiles are representative of magnetized filaments where the magnetic field influence dominates over gravity in the outer regions. For isothermal filaments, they found that $\rho(r) \propto r^{-1.8}$ to $r^{-2}$ and for logatropic filaments $\rho(r) \propto r^{-1}$ to $r^{-1.8}$. These shallower profiles may also be explained by non-isothermal, collapsing clouds, accreting the surrounding diffuse gas \citep{Nakamura1999}.

\citet{Dapp2009} and \citet{Arzoumanian2011} showed that the column density profiles of interstellar filaments can be fitted with the following radial hydrogen density distribution
\begin{equation}
\label{density_profile}
\rho(r) = \frac{\rho_C}{1 + (r/H_{0})^\alpha},
\end{equation}
where $\rho_C$ is the central density, which remains constant along the axis of symmetry, and $H_0$ is the internal flat radius. We define the ($\rho_C, H_0$)-parameters for the clouds to be at equilibrium, with their mass per unit length equal to the critical value defined by \citet{Ostriker1964}, $M_{crit} = 2 c_S^2 / G$, where $G$ is the gravitational constant. The sound speed, $c_S \sim 0.2$ km/s, is assumed to be constant and is computed for a gas temperature of $T_{gas} = 12$ K. The steepness of the density profile, $\alpha$, is usually in the range $1.5 \lesssim \alpha \lesssim 2.5$ \citep{Dapp2009, Arzoumanian2011}. For our study, we assume that $\alpha = 2$. The mass per unit length for the profile defined by Eq. \ref{density_profile} is
\begin{eqnarray}
 M &=& \int_{0}^{R} 2 \pi \rho(r) r dr \\
 M &=& \pi \rho_C H_0^2 \ln\left[ 1 + \left( \frac{R}{H_0} \right)^2 \right],
\end{eqnarray}
where $R$ is the outer radius of the cloud. Assuming that $M = M_{crit}$, the ($\rho_C, H_0$)-parameters can be solved numerically
\begin{equation}
 \pi \rho_C H_0^2 \ln\left[ 1 + \left( \frac{R}{H_0} \right)^2 \right] = \frac{2 c_S^2}{G}.
\end{equation}
We set $R = 1$ pc to be able to compare our results with those of \citet{Arzoumanian2011}, who used $R = 1.5$ pc and those of \citet{Juvela2012b}, who chose $R = 0.4$ pc. The clouds are divided into 217 cells, consisting in 31 rings and 7 layers (see Section \ref{description_CRT}). We compute the ($\rho_C, H_0$)-parameters to model clouds with visual extinction at the centre from $A_V = 1$ to 20 (Tab. 3). The emission maps produced  have $41 \times 41$ pixels with a single pixel size of 0.025 pc, equivalent to 0.85' at a distance of 100 pc or 5'' at 1 kpc. Finally, when not specified, these idealized filaments are illuminated by the isotropic interstellar standard radiation field, ISRF, as defined by \citet{Mathis1983}. We also study an alternative geometry for the clouds: Bonnor-Ebert spheres \citep{Bonnor1956, Ebert1955}. The results are presented in Appendix \ref{appendix_BE} and allow us to test the influence of the cloud geometry on the results.

\begin{table*}
\label{tableau_clouds}
\centering
\caption{Parameters defining the gas distribution of the modelled clouds. Central density ($\rho_C$ in H/cm$^3$) and internal flat radius ($H_0$ in pc) for the clouds with visual extinction at the centre $A_V = 1$ to 20, and for all the dust populations considered in the paper (see Tab. 2).}
\begin{tabular}{ccccccccccc}
\hline
\hline
\multicolumn{11}{c}{DHGL (Dust at high galactic latitude)} \\
\hline
$A_v$ & 1 & 2 & 3.5 & 5 & 7.5 & 10 & 12.5 & 15 & 17.5 & 20 \\
$\rho_C$ (H/cm$^3$) & 460 & $2\,900$ & $11\,480$ & $26\,800$ & $69\,300$ & $135\,350$ & $225\,500$ & $341\,000$ & $481\,500$ & $647\,750$ \\
$H_0$ (pc) & 0.74 & 0.16 & 0.065 & 0.039 & 0.022 & 0.015 & 0.011 & 0.0090 & 0.0075 & 0.0063 \\
\hline
\hline
\multicolumn{11}{c}{Accreted grains} \\
\hline
$\rho_C$ & 370 & $2\,300$ & $9\,350$ & $22\,000$ & $56\,800$ & $111\,000$ & $185\,600$ & $282\,000$ & $398\,000$ & $536\,500$ \\
$H_0$ & 1.04 & 0.18 & 0.073 & 0.044 & 0.025 & 0.017 & 0.013 & 0.010 & 0.0083 & 0.0070 \\
\hline
\hline
\multicolumn{11}{c}{Aggregates 0\%} \\
\hline
$\rho_C$ & 590 & $3\,620$ & $14\,100$ & $32\,800$ & $84\,500$ & $164\,500$ & $273\,800$ & $413\,000$ & $581\,800$ & $783\,000$ \\
$H_0$ & 0.55 & 0.13 & 0.057 & 0.034 & 0.020 & 0.014 & 0.010 & 0.0081 & 0.0067 & 0.0057 \\
\hline
\hline
\multicolumn{11}{c}{Aggregates 10\%} \\
\hline
$\rho_C$ & 520 & $3\,240$ & $12\,700$ & $29\,700$ & $76\,500$ & $149\,000$ & $248\,300$ & $375\,000$ & $528\,600$ & $711\,100$ \\
$H_0$ & 0.64 & 0.14 & 0.061 & 0.037 & 0.021 & 0.014 & 0.011 & 0.0086 & 0.0071 & 0.0060 \\
\hline
\hline
\multicolumn{11}{c}{Aggregates 25\%} \\
\hline
$\rho_C$ & 420 & $2\,700$ & $10\,620$ & $24\,900$ & $64\,350$ & $125\,800$ & $209\,700$ & $317\,500$ & $448\,200$ & $603\,400$ \\
$H_0$ & 0.84 & 0.16 & 0.068 & 0.041 & 0.023 & 0.016 & 0.012 & 0.0094 & 0.0078 & 0.0066 \\
\hline
\hline
\multicolumn{11}{c}{Aggregates 40\%} \\
\hline
$\rho_C$ & 340 &$2\,200$ &$8\,850$ & $20\,800$ & $54\,000$ & $105\,500$ & $176\,500$ & $268\,000$ & $378\,800$ & $510\,800$ \\
$H_0$ & 1.24 & 0.19 & 0.076 & 0.045 & 0.026 & 0.018 & 0.013 & 0.010 & 0.0085 & 0.0072 \\
\hline
\end{tabular}
\end{table*}

\subsection{Dust colour temperature and emissivity spectral index}
\label{methodes}

\begin{figure*}[!th]
\centerline{
\includegraphics[width=0.87\textwidth]{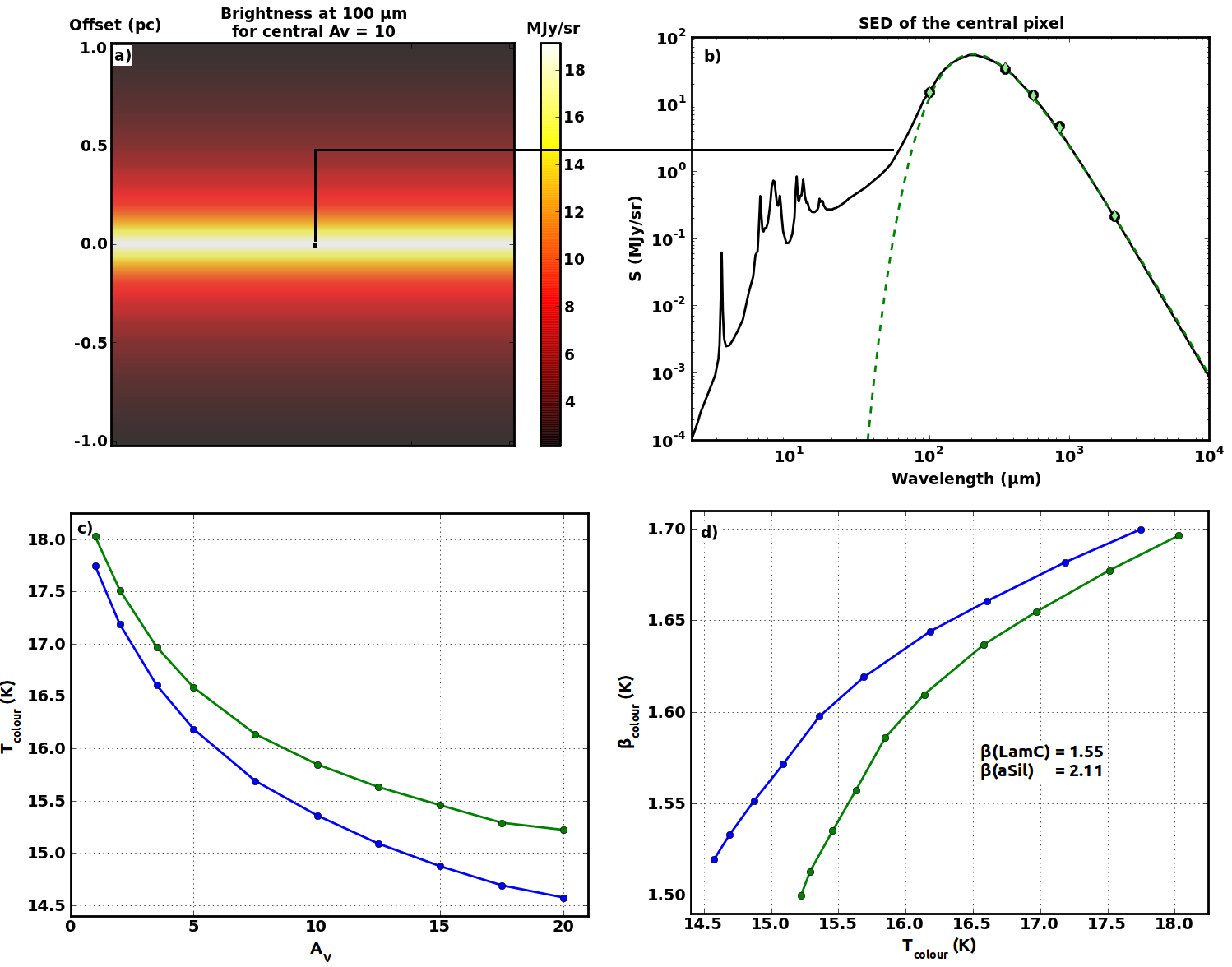}}
\caption{Case of DHGL populations. 
a) 100 $\mu$m brightness map of the cloud with $ A_V = 10$ at the centre. 
b) The black solid line is the SED of the central pixel and the black circles are the same SED integrated in Planck-HFI bands (143, 353, 545, 857 GHz) and IRIS 100 $\mu$m band. The green dashed line is the best fit between 100 $\mu$m and 3000 $\mu$m, and the green diamonds show the best fit for the SED integrated in the IRIS 100 $\mu$m and in the Planck HFI bands. 
c) Colour temperature as a function of visual extinction for the central pixel for a series of models with different central $A_V$ (see Tab. 3). The blue line shows the colour temperatures obtained by fitting the SEDs between 100 $\mu$m and 3000 $\mu$m, and the green line when the fit is made in the Planck-HFI and IRIS bands as defined in section \ref{section_diffuse}. 
d) Emissivity spectral index, $\beta_{colour}$, as a function of the colour temperature. Line styles are the same as in the previous figure. $\beta$(LamC) and $\beta$(aSil) are the intrinsic opacity spectral indices of the LamC and aSil populations. The true grain temperatures, $T_{dust}$, can be seen in Fig. \ref{Tcolour_Tdust}.}
\label{diffuse_test_case} 
\end{figure*}

Thermal emission from large dust grains dominates the observed emission in the far-IR and in the submillimetre. These grains are in thermal equilibrium with the interstellar radiation field and the observers often assume that their spectral energy distribution (SED) is well-approximated by a single grain size with a constant temperature and abundance along the line of sight
\begin{equation}
S_{\nu} = N_H \, \epsilon_0 \left(\frac{\nu}{\nu_0}\right)^{\beta_{colour}} B_{\nu}(T_{colour}),
\end{equation}
where $S_{\nu}$ is the brightness, $B_{\nu}$ is the Planck function, $T_{colour}$ is the dust colour temperature, $\epsilon_0$ is the dust emissivity at frequency $\nu_0$ (or the optical depth per unit column density $\tau_0/N_H$ in cm$^2$/H), and $\beta_{colour}$ is the emissivity spectral index. To compare our model with observations, we fit the model central pixel, which is also the brightest and the coldest pixel, with this modified blackbody where the factor $N_H \epsilon_o$, the spectral index $\beta_{colour}$, and the colour temperature $T_{colour}$ are free parameters. Three cases are considered. Except where otherwise stated, we either fit the entire SED from 100 $\mu$m to 3000 $\mu$m ($\sim$ 100 GHz), or perform the fit using the Planck-HFI\footnote{We exclude the two HFI channels which are contaminated by CO emission, 100 and 217 GHz, as in \citet{PlanckAbergel2011}.} (High Frequency Instrument) filters at 857, 545, 353, and 143 GHz (350, 550, 850, and $2\,100$ $\mu$m, respectively), and the IRIS\footnote{IRIS is the new generation of IRAS (InfraRed Astronomical Satellite) images described in \citet{Miville2005}.} 100 $\mu$m filter as in \citet{PlanckAbergel2011}, or we use the Herschel PACS\footnote{Photodetector Array Camera and Spectrometer.} and SPIRE\footnote{Spectral and Photometric Imaging Receiver.} filters at 100, 160, 250, 350, and 500 $\mu$m. Except in Section \ref{section_noise} where we consider the effects of noise, we do a standard weighted least square fits, where all bands are given an equal weight, in order to determine $T_{colour}$ and $\beta_{colour}$. The results do not depend on the absolute value adopted for the uncertainty (0.01\% here), but only on the relative weighting of the data points. The values of $T_{colour}$ and $\beta_{colour}$ are also completely dependent on the wavelength range/instrumental filters used in the fitting procedure. This dependence can be seen in Fig. \ref{diffuse_test_case}, for example, and comes from the different spectral distribution of the filters regarding the position of the peak wavelength of the big grains' emission.

\section{Dust with the properties of the diffuse high Galactic latitude ISM}
\label{section_diffuse}

\begin{figure}[!t]
\centerline{
\includegraphics[width=0.42\textwidth]{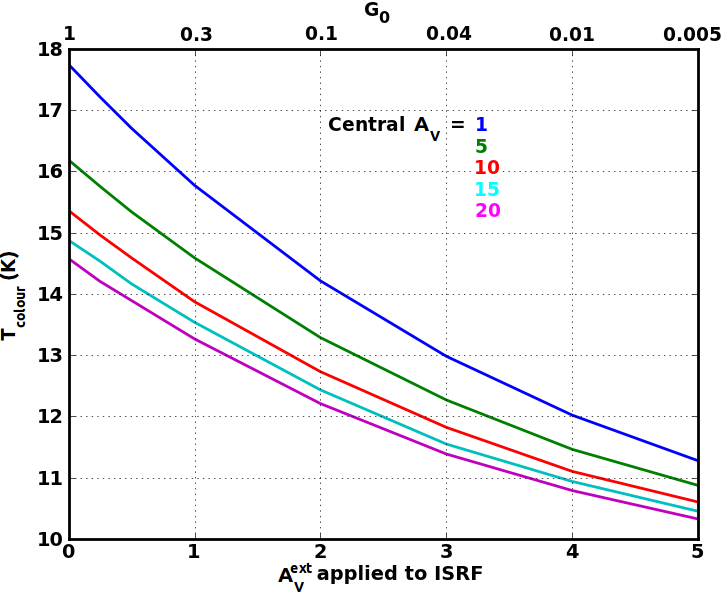}}
\caption{Colour temperature for the central brightest pixels (for SEDs fitted between 100 $\mu$m and 3000 $\mu$m) of the idealized filaments as a function of $A_V^{ext}$, which is the extinction applied to the ISRF (and the corresponding $G_0$ factor). The blue line shows the cylinder with $A_V = 1$ at the centre, green shows 5, red 10, light blue 15, and magenta 20. The true grain temperatures, $T_{dust}$, can be seen in Fig. \ref{Tcolour_Tdust}.}
\label{T_Go} 
\end{figure}

We first assume that in the ten dense idealized cylindrical filaments defined by Eq. \ref{density_profile} with the parameters listed in Tab. 3, dust has the same properties as in the diffuse ISM \citep{Compiegne2011}. The dust populations present in our modelled cylindrical clouds, with constant abundances, are thus PAHs, SamC, LamC, and aSil, described in Tab. 2 for DHGL.

\subsection{Effect of the radiative transfer}
\label{effect_of_radiative_transfer}

As shown in Fig. \ref{diffuse_test_case}c, a cloud with a visual extinction of $A_V = 1$ at the centre has a dust colour temperature higher than 17.5 K, irrespective of whether the dust SED is fitted between 100 and $3\,000$ $\mu$m or in the Planck HFI and IRIS 100 $\mu$m bands. When $A_V = 10$ or 20, the colour temperature is lower and can be as low as either 15.4 K or 14.6 K. In a cloud with high visual extinction, the radiation field at short wavelengths (visible/ultraviolet), which is mainly responsible for the grain heating, is more extinguished in the inner layers of the cloud. However, these colour temperatures are still higher by several Kelvins than what is observed in dense molecular clouds (see Tab. 1). These dense clouds are often embedded in large HI or molecular complexes, leading to a lower and reddened incident radiation field at the surface of the cloud and thus to lower colour temperatures. Using the same clouds with central $A_V = 1$ to 20, we illuminate them with an extinguished isotropic ISRF\footnote{As noted by \citet{Stamatellos2003}, the radiation field illuminating embedded clouds is usually not isotropic but we keep this assumption for the sake of simplicity.}. We extinguish the radiation by $A_V^{ext} = 0$ to 5, corresponding to $G_0 = 1$ to 0.005, respectively\footnote{$G_0$ scales the radiation field intensity integrated between 6 and 13.6 eV. The Mathis radiation field, ISRF, $G_0 = 1$, corresponds to an intensity of $1.6\times10^{-3}$ erg/s/cm$^2$.}. The results, shown for fits to the dust SEDs between 100 and $3\,000$ $\mu$m, are presented in Fig. \ref{T_Go}. A radiation field that is decreased by one magnitude of visual extinction, or with $G_0 = 0.3$, leads to a decrease in the observed colour temperature of less than 2 K. This diminution becomes smaller when the cloud central column density or visual extinction increases. Most of the UV/visible photons, which disappear when the radiation field is extinguished, have no influence on the inner layers of the densest clouds. Consequently, the extinction in the external radiation field alone cannot explain the colour temperatures measured in the objects described in Tab. 1 \citep{Bernard1992, Stepnik2003}. Similar results are obtained in the case of clouds modelled as Bonnor-Ebert spheres instead of cylinders (see Appendix \ref{appendix_BE}).

Fig. \ref{diffuse_test_case} also shows the $\beta_{colour}-T_{colour}$ relation obtained by fitting the dust SEDs between 100 and $3\,000$ $\mu$m, and in the Planck HFI and IRIS 100 $\mu$m bands. This relation exhibits a correlation instead of the observed anti-correlation between the emissivity spectral index and the colour temperature. Similar results were obtained by \citet{Malinen2011}, who modelled turbulent clouds containing gravitationally bound cores (see their Fig. 15a), and this correlation is also seen in the case of Bonnor-Ebert spheres (see Appendix \ref{appendix_BE}). It can be explained by purely radiative transfer effects for grains with constant emissivity spectral indices. The clouds are optically thick at short wavelengths and the UV/visible photons, which heat the dust grains, are absorbed efficiently in the outer regions. Thus, the dust grains are colder at the centre of the clouds than on the surface. As the clouds are optically thin at long wavelengths, we observe the emission of all the grains present along the line-of-sight. The range of temperatures increases with the central density: the dust temperature at the edge of the clouds, which are illuminated by the ISRF, remains almost the same, regardless of the central density, whereas it is significantly lower at the centre. Consequently, the resulting SED broadens as the extinction of the clouds increases, leading to a decrease in the emissivity spectral index to account for the flatter shape, while the colour temperature decreases. This $\beta_{colour}-T_{colour}$ correlation, obtained using methods including full radiative transfer calculations, contradicts the results of \citet{Shetty2009}, who found that the line-of-sight temperature variations naturally lead to a $\beta_{colour}-T_{colour}$ anti-correlation. This difference emphasizes the importance of the use of full radiative transfer calculations when modelling the dust emission from dense molecular clouds and is addressed in detail by \citet{Juvela2012}. These authors demonstrated that the anti-correlation observed by \citet{Shetty2009} results from the particular temperature structure of the objects that they modelled, which are not likely to be representative of externally heated interstellar clouds.

\subsection{Effect of noise on the $\beta-T$ relation}
\label{section_noise}

\begin{figure}[!t]
\centerline{
\includegraphics[width=0.35\textwidth]{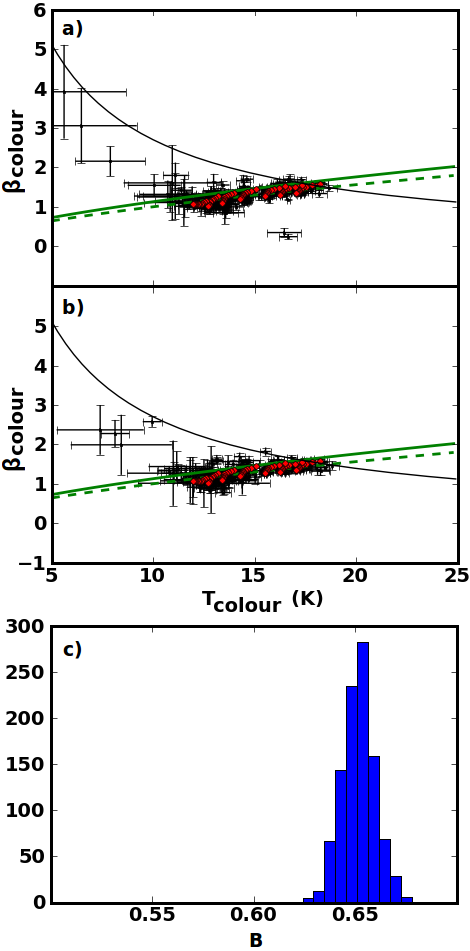}}
\caption{a) and b) Two examples of the ($T_{colour}$, $\beta_{colour}$) distributions for the observations simulated in the Planck-HFI and IRAS 100 $\mu$m bands for the central pixels of the series of filaments described in Tab. 3. The red squares show the results of the fits for the clouds described in Section \ref{effect_of_radiative_transfer}, containing DHGL grain populations and illuminated by the ISRF, extinguished or not, without noise (10 clouds $\times$ 7 radiation fields). The black crosses show the same after adding the estimated statistical noise to the simulated data (the length of the arms of the black crosses are the error bars). The green dashed lines are the power-law fits of these two particular realisations and the green solid line corresponds to the average relation over the $1\,000$ cases. For clarity, the plot includes only every 50$^{{\rm th}}$ of the $14\,000$ points (see Section \ref{section_noise} for details). The black line is the fit to the observed $\beta_{colour}-T_{colour}$ relation measured by \citet{PlanckMontier2011} towards cold clumps with Planck HFI and IRAS 100 $\mu$m bands.
c) Probability distribution of the $B$ parameter obtained from $1\,000$ cloud samples, similar to the two examples in the upper frames (with $B$ defined as $\beta_{colour} = A T_{colour}^B$, see Section \ref{section_noise} for details).}
\label{beta_T_noise_planck} 
\end{figure}

\begin{figure}[!t]
\centerline{
\includegraphics[width=0.35\textwidth]{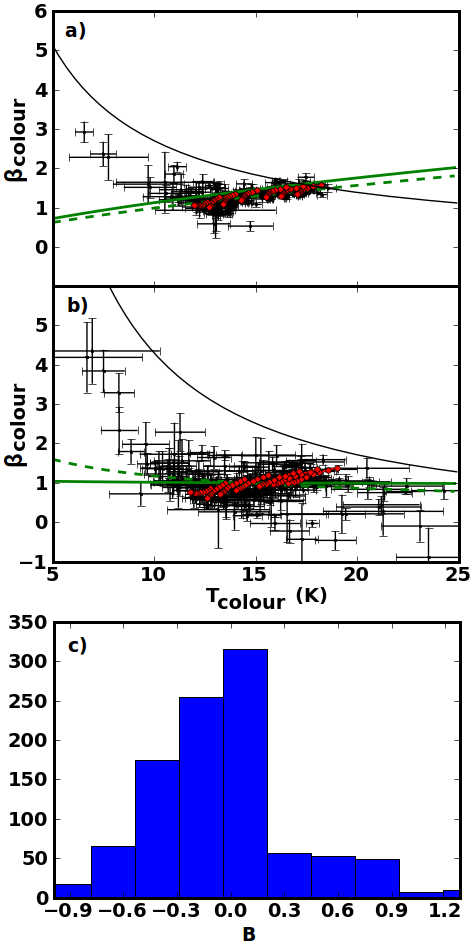}}
\caption{a), b), and c) Same as Fig. \ref{beta_T_noise_planck} for the observations simulated in the Herschel bands for the central pixels of the series of filaments described in Tab. 3. In a) and b), the black line is the fit to the observed $\beta_{colour}-T_{colour}$ relation observed by \citet{Paradis2010} in the Galactic plane with Herschel.}
\label{beta_T_noise_herschel} 
\end{figure}

Radiative transfer effects lead to a correlation between the dust emissivity spectral index, $\beta_{colour}$, and the colour temperature, $T_{colour}$, whereas observations exhibit an anti-correlation (see Tab. 1). \citet{Shetty2009} and \citet{Blain2003} showed that this anti-correlation can be produced by intrinsic noise in the observations, which produces a degeneracy between $T_{colour}$ and $\beta_{colour}$. This anti-correlation has been observed in various interstellar environments by Planck and Herschel instruments, for instance by \citet{PlanckAbergel2011, PlanckMontier2011}, \citet{Paradis2010}, \citet{Rodon2010}, \citet{Juvela2011} and \citet{Bracco2011}. These authors suggest that the observed $\beta_{colour}-T_{colour}$ anti-correlation could be produced mostly by intrinsic variations in the dust optical properties with temperature rather than noise.

To estimate qualitatively the extent to which noise could lead to an apparent $\beta_{colour}-T_{colour}$ anti-correlation, we examine the effect of observational noise on the $(\beta_{colour}-T_{colour})$ relation with the help of Monte Carlo simulations. We consider the ten clouds described in Tab. 3, assuming that they are all located at a distance of 100 pc, with DHGL grain populations, and we use the associated surface brightness maps for different external radiation fields: ISRF and six values of attenuation of the external radiation field corresponding to the ISRF extinguished by $A_V^{ext}$ up to 5 (Fig. \ref{T_Go}).

We compute $1\,000$ realisations of the observations of the modelled clouds. In each realisation, the surface brightness values are modified by the addition of normally distributed observational noise. We consider the case of observations obtained with the combination of Planck-HFI and IRAS 100 $\mu$m bands, for which the modelled brightness maps are smoothed to a common resolution of 5'. We also study the case of observations made with the Herschel PACS and SPIRE instruments for a common resolution of 37''. The standard deviation in the noise is assumed to be 0.06, 0.12, 0.12, and 0.08 MJy/sr in the IRAS 100 $\mu$m band and in the Planck HFI bands at 857, 545, and 353 GHz, respectively \citep{PlanckBernard2011}. In the case of Herschel, absolute values of noise equal to 8.1, 3.7, 1.2, 0.85, and 0.35 MJy/sr (at the beam resolution) are used for the five bands at 100, 160, 250, 350, and 500\,$\mu$m, respectively \citep{Malinen2011}. For each realisation (observations of all frequencies, including the noise), the $T_{colour}$ and $\beta_{colour}$ values are estimated using a standard $\chi^2$ method for the central pixels of the series of filaments described in Tab. 3. Their uncertainties come from Monte Carlo simulations. These results are used to simulate surveys that would equally cover clouds from all the examined cloud categories (with $A_V = 1$ to 20 and variable attenuation of the ISRF $A_V^{ext} = 0$ to 5). For each modelled cloud, we extract 200 random realisations, i.e., 200 $T_{colour}$ and $\beta_{colour}$ values derived from observations of that model with 200 different Monte Carlo estimates of the observational noise. This leads to a sample of $14\,000$ ($T_{colour}$, $\beta_{colour}$) points (10 clouds $\times$ 7 radiation field intensities $\times$ 200 noise realisations). Figures \ref{beta_T_noise_planck}a and b, and Figs. \ref{beta_T_noise_herschel}a and b show four examples of these realisations, where for clarity only every 50$^{th}$ of the $14\,000$ points are plotted. The red dots show the results obtained when noise is not taken into account: for comparison, the green line of Fig. \ref{diffuse_test_case}d, obtained for the ten modelled clouds illuminated by the ISRF, corresponds to the red dots that are at the extreme right of Figs. \ref{beta_T_noise_planck}a and b. Then moving to the left, the red dots correspond to decreasing intensities of the radiation field or similarly to increasing $A_V^{ext}$. We then proceed to fit the $14\,000$ ($T_{colour}$, $\beta_{colour}$) points with a relation $\beta_{colour}=A T_{colour}^{B}$, which is commonly used to fit real observations, using a standard $\chi^2$ method, where the points are weighted according to their error estimates\footnote{In Figs. \ref{beta_T_noise_planck}a and b, and Figs. \ref{beta_T_noise_herschel}a and b, the outliers with low colour temperatures and high emissivity spectral indices are not captured by the fits (green lines). These points indeed represent fewer than 3\% of the total number of points in the sample in the case of Planck, and fewer than 5\% in the case of Herschel, and have large error bars.}. The sign of the $B$ parameter indicates whether the $\beta_{colour}-T_{colour}$ relation is correlated ($B > 0$) or anti-correlated ($B < 0$). These procedures, the random selection of the $14\,000$ points and the fitting of the power-law, are repeated $1\,000$ times to get information about the distributions of the $A$ and $B$ parameters (Figs. \ref{beta_T_noise_planck}c and \ref{beta_T_noise_herschel}c).

In the case of Planck (+IRAS), Fig. \ref{beta_T_noise_planck}c shows that $0.62 \leqslant B \leqslant 0.68$. Even if flattened, the relation between $\beta_{colour}$ and $T_{colour}$ remains positively correlated with $\beta_{colour} \propto T_{colour}^{0.65}$. In the case of Herschel, the values of $B$ are more scattered. However, steep relations with $B\le-1$ are excluded at a 99.9\% confidence level, while relations steeper than $B=-0.5$ are found for only $\sim$ 10\% of the cloud samples. The difference between the Planck and Herschel cases comes from the high level of noise introduced in the PACS channels.

\citet{PlanckMontier2011} used a combination of Planck-HFI data (857, 545, and 353 GHz) and IRIS 100 $\mu$m data to determine the dust properties in the starless cold clumps described in the Cold Core Catalogue of Planck Objects (C3PO). They observed an anti-correlation that could be fitted with $\beta_{colour} \sim T_{colour}^{-1}$. \citet{PlanckAbergel2011} found a similar $\beta_{colour}-T_{colour}$ relation in the Taurus molecular complex, using Planck-HFI and IRAS bands. In the case of Herschel, \citet{Paradis2010} measured an anti-correlation that could be fitted with $\beta_{colour} \sim T_{colour}^{-1.3}$ in the Galactic plane, while \citet{Bracco2011} found a comparable relation in a Galactic cirrus. Consequently, if we assume that our sample of modelled clouds is representative of the ranges of dust temperature and opacity spectral index of the observations, noise as the unique explanation of the observed anti-correlation appears to be unlikely. To test the influence of the cloud geometry, we did the same analysis using externally heated Bonnor-Ebert spheres and found similar results (see Appendix \ref{appendix_BE}).

As pointed by \citet{Shetty2009} and \citet{Blain2003}, intrinsic noise in the data produces a degeneracy between $\beta_{colour}$ and $T_{colour}$ potentially leading to an anti-correlation (Fig. \ref{beta_T_noise_herschel}) or at least a flattening of the relation (Fig. \ref{beta_T_noise_planck}). Nevertheless, in our simulations noise alone cannot explain the strong anti-correlation observed. For starless interstellar clouds, it seems probable that the anti-correlation is explained instead by variations in the dust optical properties with some contribution from the noise (see Section \ref{section_beta_T}). However, if the observers do not exclude sources with very low signal-to-noise ratios (S/N's) (e.g., $S/N<4$ at 250 $\mu$m, see Appendix B), the noise can then have a dominant role on the observed $\beta_{colour}-T_{colour}$ relation.

\section{Evolution of grains with medium density: accretion and aggregation}
\label{section_grain_growth}

The size of the dust grains is expected to increase with the density of the medium \citep{Ossenkopf1993, Bernard1999, Stepnik2003, Steinacker2010}, hence we investigate the effects of grain growth on the long-wavelength SED of dense clouds. The parameters we consider when comparing with observations are the emission deficit in the mid-IR, the higher emissivity in the submillimetre, the decrease in the grain colour temperature, and the $\beta_{colour}-T_{colour}$ relation. The deficit in the mid-IR is usually explained by a lower abundance of small stochastically heated grains (PAH and SamC populations), and the higher submillimetre emissivity by coagulation \citep{Ossenkopf1994, Stognienko1995}. Four scenarii for grain growth can be considered:
\begin{itemize}
 \item aggregation of small grains with small grains ;
 \item aggregation of big grains with big grains ;
 \item accretion of small grains onto big grains (hereafter accretion scenario) ;
 \item aggregation of all grains with all grains (hereafter aggregation scenario).
\end{itemize}
The first scenario, aggregation of small grains with small grains, allows us to reproduce the emission deficit in the mid-IR but has no significant impact on the grain submillimetre emissivity: large grain submillimetre emission and colour temperature are almost unchanged in this scenario. The second scenario leads to a decrease in the big grain colour temperature and an increase in the submillimetre emissivity, but does not explain the emission deficit in the mid-IR. The two last scenarii both explain qualitatively the emission deficit in the mid-IR with the loss of all the small grains. We now describe in detail their effects on the dense cloud SEDs. To perform this analysis, we consider that the ten cylindrical clouds described in Tab. 3 contain only either accreted grains (Section \ref{accretion}) or aggregates (Section \ref{coagulation}).

\subsection{Accretion}
\label{accretion}

\begin{figure}[!t]
\centerline{
\includegraphics[width=0.42\textwidth]{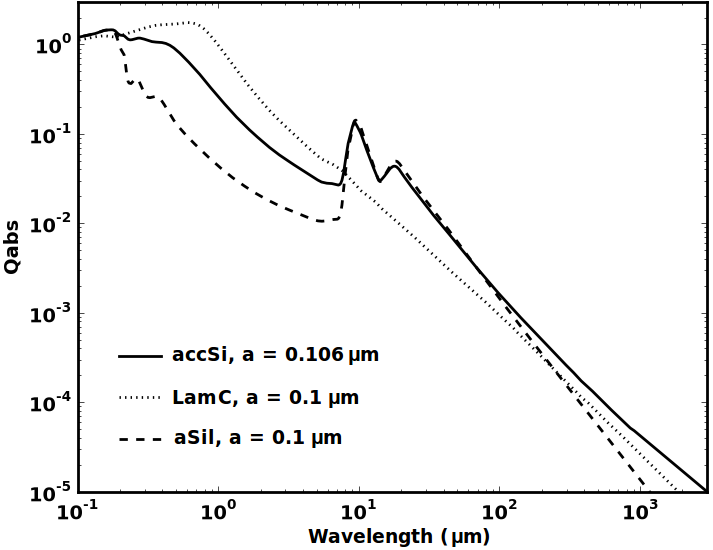}}
\caption{Absorption efficiencies, $Q_{abs}$, for an aSil grain (dashed line), a LamC grain (dotted line), and an accSi grain (solid line), with sizes as defined in section \ref{accretion}.}
\label{Qabs_accretion} 
\end{figure}

All the small carbonaceaous particles (PAHs and SamC) are accreted at the surface of larger grains (aSil and LamC) as uniform carbon mantles of the same thickness, regardless of the size considered, to form accreted carbonaceous particles and accreted silicates (accC and accSi, see Tab. 2.) This increases the total mass of large grains by 10\% and their volume by 17\%, while their radius remains almost unchanged ($\sim +6$\%).

The LamC and accC populations have the same optical properties and size distribution. In contrast, accreted silicates, accSi, are composed of 84\% of aSil and 16\% of carbon in volume (the amC mantle has a specific density of 1.81 g/cm$^3$, compared with 3.5 g/cm$^3$ for silicates, see Tab. 2). Their optical properties are calculated with the effective medium theory (EMT), using the Bruggeman rule \citep{Bohren1983}. The dielectric function of the composite medium, $\epsilon_{\rm eff}$, is the solution to Eq. 6 involving the dielectric function of the first component, $\epsilon_1$, of fractional volume $f$, and the second component, $\epsilon_2$, of fractional volume $1-f$ given by
\begin{equation}
f\,\frac{\epsilon_1 - \epsilon_{\rm eff}}{\epsilon_1 + 2\epsilon_{\rm eff}} + (1-f)\,\frac{\epsilon_2 - \epsilon_{\rm eff}}{\epsilon_2 + 2\epsilon_{\rm eff}} = 0.
\end{equation}
For a two-component mix similar to that considered here, silicateous and carbonaceous, this equation is simply quadratic in $\epsilon_{\rm eff}$.

The variation in the optical properties of the accreted grains compared with aSil and LamC are shown in Fig. \ref{Qabs_accretion}, for grains containing the same amount of silicate ($a_{{\rm aSil}} = 0.1 \; \mu{\rm m} \Leftrightarrow a_{{\rm accSi}} = 0.106 \; \mu{\rm m}$). The absorption efficiency of accSi grains, $Q_{abs}$, in the far-IR and the submillimetre is flatter ($\beta = 1.52$) than for aSil ($\beta = 2.11$) and LamC ($\beta = 1.55$): at 250 $\mu$m, it is multiplied by a factor of 1.54. $Q_{abs}$ also increases in the visible and in the near-IR. As a result, accSi grains absorb more energy than aSil grains. Both their submillimetre emissivity and their temperature are higher than for DHGL dust (Fig. \ref{grain_growth}a). Despite being enhanced, their emissivity at long wavelengths is indeed not high enough to counterbalance the larger amount of energy absorbed at shorter wavelengths. This increase in the colour temperature does not match the observations. We also note that the accSi grains have a lower spectral index value that is, however, correlated with the colour temperature (Fig. \ref{grain_growth}b). The results are qualitatively similar for Bonnor-Ebert spheres (see Appendix \ref{appendix_BE}).

\subsection{Aggregation with or without porosity}
\label{coagulation}

\begin{figure}[!t]
\centerline{
\includegraphics[width=0.42\textwidth]{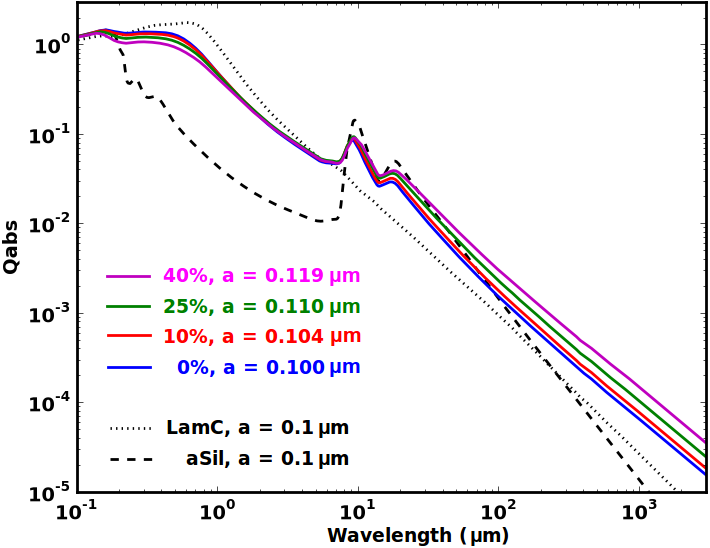}}
\caption{Absorption efficiencies, $Q_{abs}$, for an aSil grain (black dashed line), for a LamC grain (dotted black line), for an aggregate without porosity (blue), with a porosity degree of 10\% (red), 25\% (green), and 40\% (magenta), with radii as defined in section \ref{coagulation}. All these grains have the same volume of material, excluding the void only responsible for the increase of the size.}
\label{Qabs_coagulation} 
\end{figure}

In this scenario, all grains are aggregated together to form mixed aggregates of small and big grains, which are necessarily porous. These aggregates are expected to form in dense molecular clouds, possibly with large fractions of void or porosity \citep{Dorschner1995}. In our simple model, the fraction of silicate, carbon, and vacuum in the grains do not depend on their sizes. It is defined using the fractional porosity parameter $p$. The aggregates are composed of $63(1-p)$\% silicate, $37(1-p)$\% carbon, and $100p$\% vacuum. These proportions ​​correspond to the conservation of the mass during the aggregation of the DHGL grain populations defined in Tab. 2. As the total volume of big grains is dominated by silicates, the size distribution is modified starting from the aSil size distribution. Grains are on average larger and the size distribution is multiplied by a factor $(0.63\times(1-p))^{-1/3}$. From $a^{-3.4}$, the power-law then becomes $a^{-2.4}$ to mimic the loss of the small grains by aggregation, which allows us to reproduce the trend of an increase in $R_V$ towards dense media \citep{Fitzpatrick1988, Mathis1989, Cardelli1991, Campeggio2007}.

The optical properties of this new grain population are computed with the Bruggeman rule, as presented in Section \ref{accretion}, but with three components of silicate, carbon, and void ($\epsilon = 1$). We first mix the carbon and silicate materials as in Section \ref{accretion},with their respective volume fractions of 63\% and 37\%. We then mix the resulting grains with void assuming the same two-component Bruggeman rule. This theory of an effective medium does reproduce the general {\it trend} that we want to test here, where the inclusion of porosity is a way to increase the submillimetre emissivity of grains. For a precise calculation of the optical properties of composite aggregates, heavy computational methods such as DDA or T-MATRX must be used \citep{Stognienko1995, Fogel1998, Mackowski2006, Kohler2011}.
 
\begin{figure}[!t]
\centerline{
\includegraphics[width=0.42\textwidth]{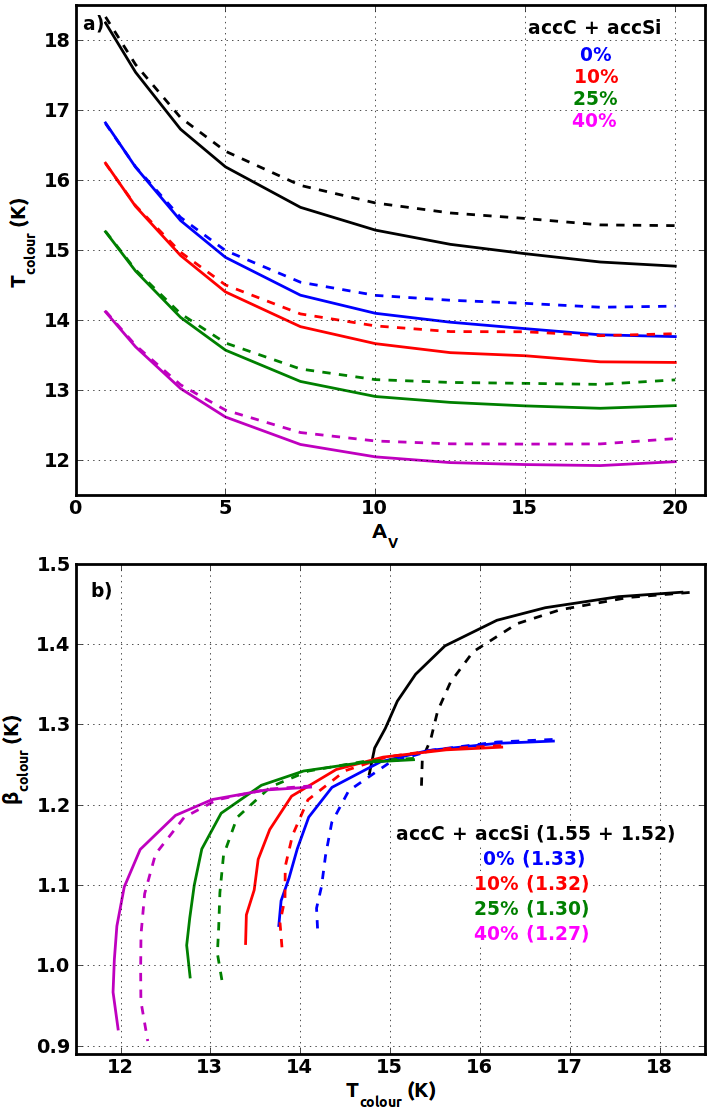}}
\caption{a) Colour temperature as a function of visual extinction for the central pixel. Results are shown for accreted and aggregated grains as defined in Tab. 2. Solid lines show the colour temperature calculated by fitting the SEDs between 100 $\mu$m and 3000 $\mu$m, and the dashed lines in the Planck-HFI bands as defined in section \ref{methodes}. The black line shows the case of accreted grains, the blue line of aggregates without porosity, red of a porosity degree of 10\%, green 25\%, and magenta 40\%, as defined in Tab. 2. The true grain temperatures, $T_{dust}$, can be seen in Fig. \ref{Tcolour_Tdust}.
b) Emissivity spectral index as a function of colour temperature. Line styles are the same as in the top figure. The numbers in parenthesis are the intrinsic opacity spectral indices of the different grain populations.}
\label{grain_growth} 
\end{figure}

The variation in the optical properties of the different populations of aggregates compared with those for the aSil and LamC grains are shown in Fig. \ref{Qabs_coagulation}. We compare grains containing the same volume of material, meaning that the void is excluded from the calculation of the volume for the aggregates. The aggregation of carbonaceous and silicate grains increases the absorption efficiency in the near-IR and the visible with respect to pure silicate cores. The trend is similar in the submillimetre: the submillimetre emissivity is increased first by the mixing between carbons and silicates, and second by the addition of porosity, a well-known effect \citep{Voshchinnikov2006}. The increase in the emissivity for the aggregates wins over the increase in absorptivity, so that the colour temperature of the aggregates is lower and decreases with increasing porosity (Fig. \ref{grain_growth}a). In addition, the emissivity spectral index in the submillimetre decreases from $\beta = 2.11$ for aSil, to 1.33 for aggregates without porosity, 1.32 with 10\% porosity, 1.30 with 25\% porosity, and 1.27 with 40\% porosity. Through aggregation, the dust emissivity is increased by a factor equal to the ratio of the $Q_{abs}$ of the aggregates to the mean $Q_{abs}$ of isolated DHGL carbon and silicate grains of the same volume. As carbon and silicate 0.1 $\mu$m grains have almost the same $Q_{abs}$ at 250 $\mu$m (Fig. \ref{Qabs_accretion}), the increase in the dust emissivity brought about by aggregation is simply the ratio of the $Q_{abs}$ of the aggregates of Fig. \ref{Qabs_coagulation} to the $Q_{abs}$ of a 0.1 $\mu$m silicate (or carbon) grain. We find an increase of 1.84 for a grain without porosity, 2.16 for a grain with 10\% of porosity, 2.84 for 25\%, and 3.89 for 40\%. When the modification of the size distribution is taken into account, these factors are even higher: in the case of DHGL populations, only 83\% of the total volume of grains emit in the submillimetre (excluding PAHs and SamC), whereas 100\% do in the case of aggregates (PAHs and SamC are now aggregated with big grains). Thus, the factors become 2.2, 2.5, 3.3, and 4.6 for aggregates with a porosity fraction of 0, 10, 25, and 40\% respectively. For a cloud with a central $A_V = 10$, this leads to a higher "observed" emissivity by a factor of 3.1, 3.6, 4.8, and 6.9, for grains with a porosity degree of 0, 10, 25, and 40\% respectively, when the dust SEDs are fitted in Planck bands. These values agree well with what is observed in dense molecular clouds (see Tab. 1). We also note that the "observed" emissivity spectral index, $\beta_{colour}$, is again correlated with the colour temperature in an opposite way to what is observed in the ISM (Fig. \ref{grain_growth}b). Similar results are obtained for Bonnor-Ebert spheres (see Appendix \ref{appendix_BE}).

The aggregation of grains in mixed aggregates of small and big grains is able to explain simultaneously the deficit of emission in the mid-IR, the low colour temperature, and the higher submillimetre emissivity observed in dense molecular clouds (see Tab. 1). However, grain growth cannot reproduce the $\beta_{colour}-T_{colour}$ anti-correlation, which might originate in grain intrinsic properties and in particular in variations in the emissivity spectral index with the wavelength and/or the temperature $\beta(\lambda, T)$ (see Section \ref{dust_intrinsic_properties}).

\subsection{Colour temperature versus grain temperature}
\label{section_grain_temperature}

\begin{figure*}[!th]
\centerline{
\includegraphics[width=0.8\textwidth]{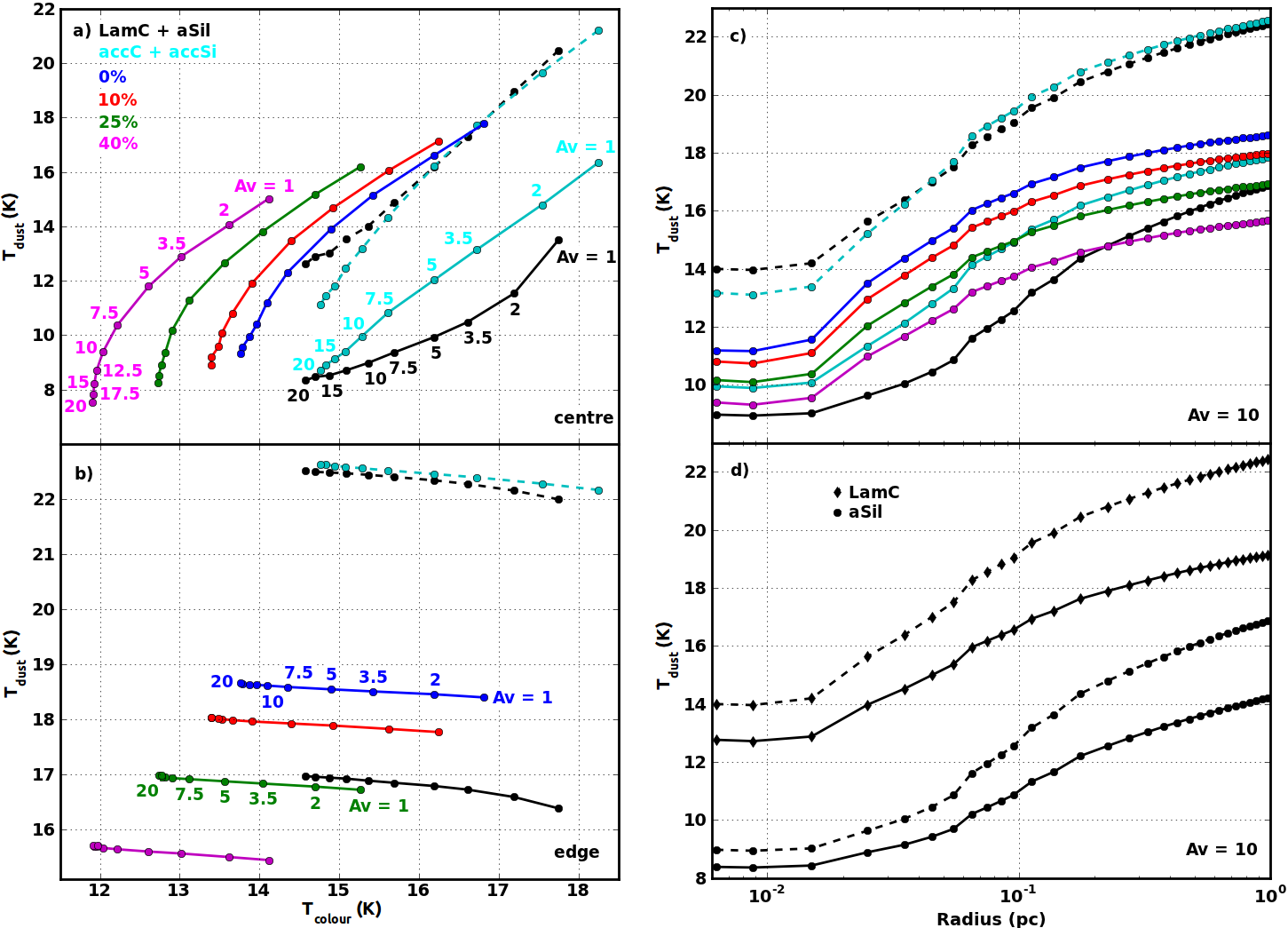}}
\caption{a) Equilibrium temperature of dust in the central cell of the modelled clouds, $T_{dust}$, averaged over the grain size distribution, as a function of the colour temperature, $T_{colour}$, calculated by fitting the SEDs between 100 $\mu$m and 3000 $\mu$m. The DHGL case is represented by the black lines (the solid line is aSil and the dashed line is LamC), the accretion case by the cyan lines (the solid line is accSi and the dashed line is accC), the coagulation case with 0\% of porosity by the blue line, with 10\% of porosity by the red line, with 25\% by the green line, and with 40\% by the magenta line.
b) As in Fig. \ref{Tcolour_Tdust}a) but at the edge of the cloud.
c) $T_{dust}$ as a function of the cloud radius for a cloud with central $A_V = 10$. The colour code is the same as in \ref{Tcolour_Tdust}a).
d) The same for the DHGL case (the circles show aSil and the diamonds LamC) for a cloud illuminated by the standard ISRF (dashed line) and illuminated by the ISRF extinguished by one visual magnitude (solid line).}
\label{Tcolour_Tdust} 
\end{figure*}

In dense molecular clouds, the intensity of the radiation field decreases from the edge to the centre. Consequently, dust grains have different temperatures for different positions inside the cloud and these temperatures differ from the measured colour temperature. Figure \ref{Tcolour_Tdust} shows the relation between the colour temperature, $T_{colour}$, and the temperature of the grains, $T_{dust}$, which is defined as the individual equilibrium temperatures averaged over the grain size distribution. We determine one dust temperature per cell and per grain population.

Fig. \ref{Tcolour_Tdust}a presents the dust temperature at the centre of the clouds as a function of the colour temperature obtained by fitting the dust SEDs between 100 and $3\,000\;\mu$m. Except for the purely carbonaceous grains (LamC), $T_{dust}$ is always lower than $T_{colour}$ for clouds with $A_V > 1$. This is the result of the mixing of all the dust temperatures from the warmer edge to the colder centre of the clouds. The relation between the central $T_{dust}$ and $T_{colour}$ is almost linear for the clouds with central $A_V \leqslant 10$. This means that for this kind of moderately dense clouds we should be able to estimate the physical dust temperature at the centre, assuming one dust population and knowing the colour temperature. The results of these linear fits are given in Tab. 4. However, we expect these relations to be more complicated if several dust populations are mixed along the line-of-sight, for instance DHGL grains at the edge and aggregates at the centre if we assume that the grains evolve within dense molecular clouds. Figure \ref{Tcolour_Tdust}b) shows the dust temperature at the edge of the clouds as a function of the colour temperature. For $A_V = 1$ to 20, $T_{dust}$ does not vary by more than 0.5 K and, as expected, its values remain close to the dust temperature measured for the same population of grains illuminated by the ISRF, without radiative transfer. However, we note a slight increase in $T_{dust}$ when $T_{colour}$ decreases, or similarly when the central $A_V$ increases. For variations in $T_{colour}$ from about 2.5 K to 3.5 K, depending on the dust population considered, $T_{dust}$ increases by $\sim 0.5$ K. This surprising rise in the dust temperature can be explained by the decrease in the central flat radius, $H_0$, with increasing central density, $\rho_C$, leading to a slightly higher incident radiation field in the outer layers (Tab. 3).

Fig. \ref{Tcolour_Tdust}c displays the dust temperature as a function of the radial position in the cloud, for a cloud with a visual extinction at the centre equal to 10. The temperature depends on the grain composition and it decreases from 7 K to 9.5 K from the edge to centre, in agreement with previous studies \citep{Bernard1992, Evans2001, Zucconi2001, Fischera2011}. For the central layers of the cloud, the dust temperature is almost constant and equal to from $\sim 8.5$ K to 14 K depending on the dust population, the hottest grains being the carbonaceous ones (LamC). This is explained by grains being mostly heated by the visible and UV photons of the incident radiation field, which are mainly absorbed in the outer layers of the cloud. This behaviour is found for all the clouds we modelled and is the most visible when the central $A_V$ is the highest. We also note that the decrease in the dust temperature from the edge to the centre of the cloud is smaller when the incident radiation field is extinguished (Fig. \ref{Tcolour_Tdust}d), as similarly observed by \citet{Stamatellos2003} and \citet{Fischera2011}. The more the radiation field is extinguished, the smaller the temperature variation.

\begin{table}
\label{linear_fits}
\centering
\caption{Linear fits between the dust temperature at the centre of the clouds as a function of their colour temperature and their dust content, for clouds with a central $A_V \leqslant 10$.}
\begin{tabular}{lcl}
\hline
\hline
\multicolumn{2}{c}{Dust populations} & Linear fits for $A_V \leqslant 10$ \\
\hline
Big grains (DHGL) & LamC & $T_{dust} \sim 2.71 T_{colour} - 27.68$ \\
                  & aSil & $T_{dust} \sim 1.79 T_{colour} - 18.89$ \\
\hline
Accreted grains & accC  & $T_{dust} \sim 2.70 T_{colour} - 27.73$ \\
                & accSi & $T_{dust} \sim 2.12 T_{colour} - 22.36$ \\
\hline
Aggregates & 0\%  & $T_{dust} \sim 2.36 T_{colour} - 21.56$ \\
           & 10\% & $T_{dust} \sim 2.39 T_{colour} - 21.35$ \\
           & 25\% & $T_{dust} \sim 2.47 T_{colour} - 21.23$ \\
           & 40\% & $T_{dust} \sim 2.62 T_{colour} - 21.65$ \\
\hline
\end{tabular}
\end{table}

\section{Dust intrinsic properties}
\label{dust_intrinsic_properties}

Grain growth and purely radiative transfer effects can explain neither the $\beta_{colour}-T_{colour}$ anti-correlation, nor the variations in the emissivity spectral index with wavelength that are observed in the ISM. These two observations may thus originate in variations in the grain intrinsic properties. Hence, we study the effects of variations in the optical properties of the grains with wavelength, $\beta(\lambda)$, and then temperature, $\beta(T_{dust})$, on the dust SEDs of the clouds.

\subsection{Variations in optical properties with wavelength: $\beta(\lambda)$}
\label{spectral_variations}

\begin{figure}[!t]
\centerline{
\includegraphics[width=0.42\textwidth]{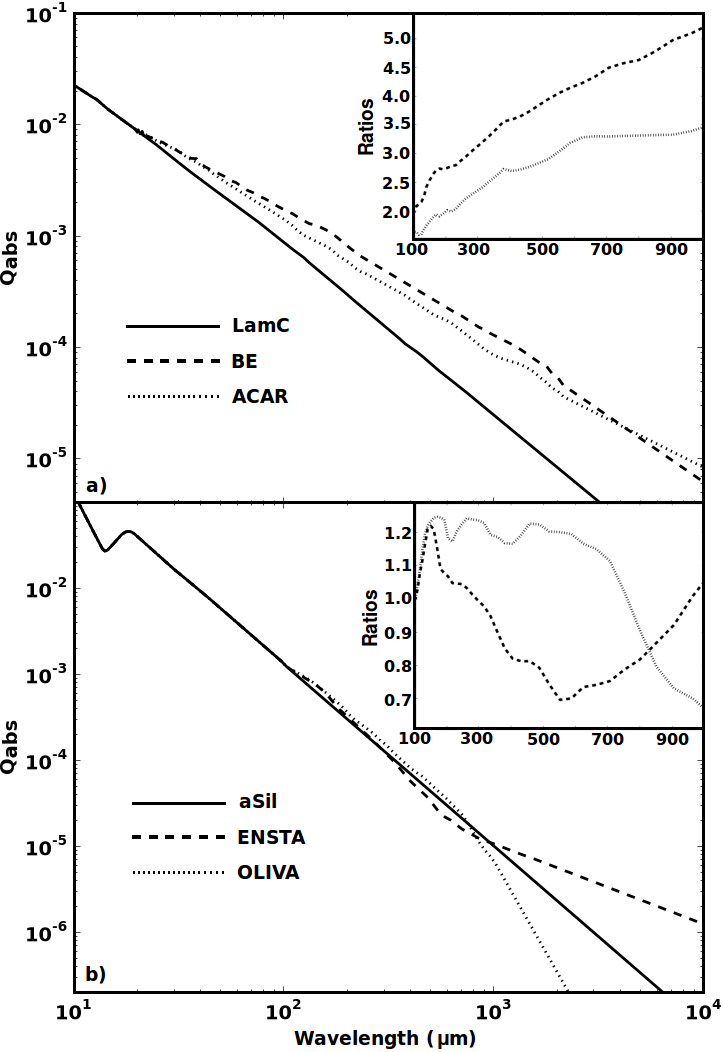}}
\caption{a) Absorption efficiencies for LamC (solid line), BE (dashed line), and ACAR (dotted line), for grains with $a = 0.1$ $\mu$m. The small window shows the ratios of the absorption efficiencies of BE and ACAR to LamC with the same linestyles.
b) Absorption efficiencies for aSil (solid line), ENSTA (dashed line), and OLIVA (dotted line), for grains with $a = 0.1$ $\mu$m. The small window shows the ratios of the absorption efficiencies of ENSTA and OLIVA to aSil with the same linestyles.}
\label{beta_lambda_1} 
\end{figure}

\begin{figure}[!t]
\centerline{
\includegraphics[width=0.42\textwidth]{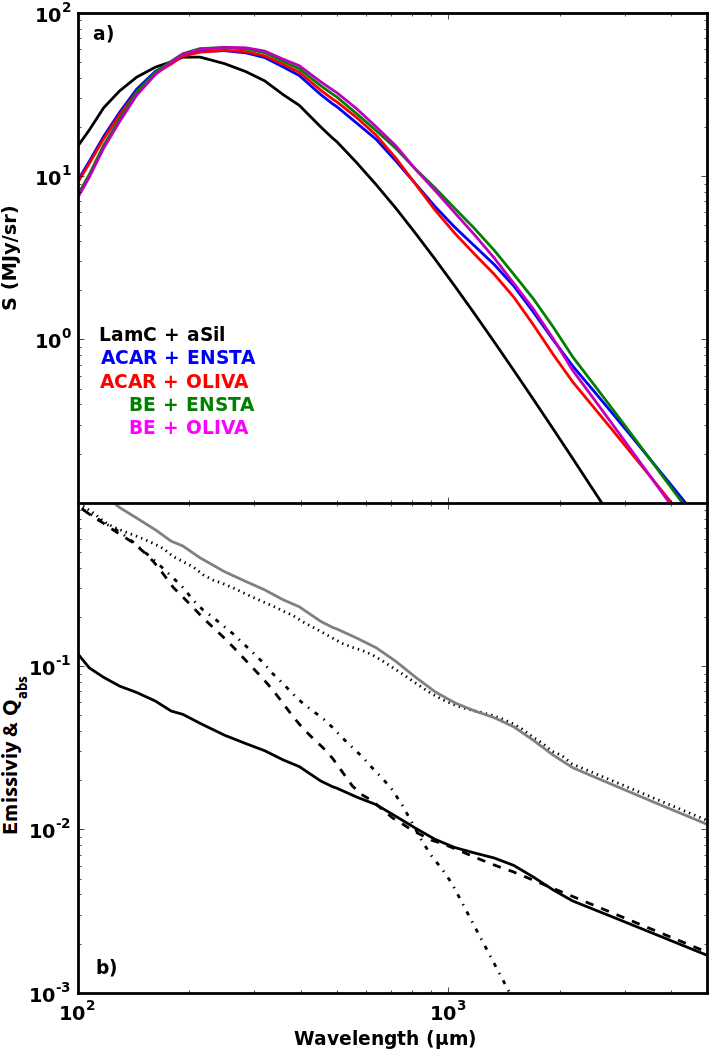}}
\caption{a) SEDs for the central pixel of the cloud with central $A_V = 10$ for a mixture of LamC and aSil (black), BE and ENSTA (green), BE and OLIVA (magenta), ACAR and ENSTA (blue), and ACAR and OLIVA (red).
b) Emissivity in arbitrary units for the cloud with central $A_V = 10$, for a mixture of ACAR and OLIVA (solid grey line), and a mixture of ACAR and ENSTA (solid black line). Absorption efficiencies in arbitrary units for grains with a diameter of 0.1 $\mu$m: ACAR (dotted line), ENSTA (dashed line), and OLIVA (dashed-dotted line).}
\label{beta_lambda_2} 
\end{figure}

We investigate the impact of spectral variations in the grain optical properties on the emerging emission from dense molecular clouds. We perform this analysis using laboratory data in the FIR and the submillimetre to modify the absorption efficiencies, $Q_{abs}$, of LamC and aSil grains. We consider samples of amorphous materials, which are representative of interstellar grains \citep{DD2003, Kemper2004, Fitzpatrick2007}.

\citet{Mennella1998} measured the absorption coefficient of two different amorphous carbonaceous cosmic analogue grains between 20 $\mu$m and 2 mm at 24 K. Their first sample was composed of a population of chainlike aggregates of spheroidal grains with a diameter around 10 nm (ACAR sample), and did not contain hydrogen. The second sample consisted of a soot with the same morphology but for spheroidal grains with a diameter of around 30 nm (BE sample). The BE sample is representative of $sp^2$-rich hydrogenated amorphous carbon. To build the absorption cross-sections of the two dust populations, for wavelengths shorter than 20 $\mu$m, we use the absorption efficiency of the LamC population and normalise the ACAR and BE absorption coefficients at 20 $\mu$m for longer wavelengths. For wavelengths longer than 2 mm, we then extrapolate $Q_{abs}$ using a unique spectral index as measured at 2 mm: $\beta(2 \; {\rm mm}) = 0.95$ for the ACAR population and 1.30 for the BE population. The results are shown in Fig. \ref{beta_lambda_1}a). Both the ACAR and BE populations have a flatter $Q_{abs}$ than the LamC population in the far-IR and in the submillimetre (for $\lambda > 20 \;\mu$m). At 250 $\mu$m, when compared with LamC, the absorption efficiency is higher by a factor of 2.2 for an ACAR grain with $a = 0.1$ $\mu$m and by 2.9 for a BE grain with the same size. The BE grains have a higher absorption efficiency than the ACAR grains for wavelengths between 20 and 3000 $\mu$m. For the ACAR and BE populations, we use the same size distribution and abundance as used for LamC grains.

For the silicates, we use the data from \citet{Coupeaud2011}, who measured the opacity of interstellar dust analogues at 10 K between 100 $\mu$m and 1500 $\mu$m. Two of their samples are considered. The first is composed of amorphous silicate with a stoichiometry close to olivine (Mg$_{2.3}$SiO$_4$, sample F1 in Coupeaud's nomenclature, OLIVA hereafter) and the second of amorphous silicate with a stoichiometry close to enstatite (MgSiO$_3$, sample E, hereafter ENSTA). In both cases, for wavelengths shorter than 100 $\mu$m, we use the absorption efficiency of aSil and for longer wavelengths, we normalise the OLIVA and ENSTA absorption coefficients at 100 $\mu$m to produce the absorption efficiencies. Finally, for wavelengths longer than $1\,500$ $\mu$m, we extrapolate $Q_{abs}$ using a unique spectral index as measured at $1\,500$ $\mu$m: $\beta($1\,500$ \; \mu{\rm m}) = 4.30$ for OLIVA and 0.95 for ENSTA. The results are shown in Fig. \ref{beta_lambda_1}b. The emissivity spectral index of the absorption efficiency of OLIVA changes from $\beta = 2.1$ below $\sim$ 700 $\mu$m to $\beta=3.6$ above, and the spectral index of ENSTA from $\beta = 2.5$ for $\lambda \leqslant 600 \; \mu$m, to $\beta = 1.7$ for $600 \leqslant \lambda \leqslant 800 \; \mu$m, and to $\beta = 0.9$ for longer wavelengths (Coupeaud et al. 2011, see their Tab. 3). These indices deviate from the aSil emissivity spectral index, which is equal to 2.11 for $\lambda \geqslant 100 \; \mu$m (Fig. \ref{beta_lambda_1}b). For wavelengths longer than $1\,200/1\,500 \; \mu$m, the absorption efficiency of OLIVA/ENSTA is extrapolated and the difference from aSil, although very pronounced, is questionable. However, this wavelength range does not influence the results of the models (see the following paragraph).

\begin{figure}[!t]
\centerline{
\includegraphics[width=0.42\textwidth]{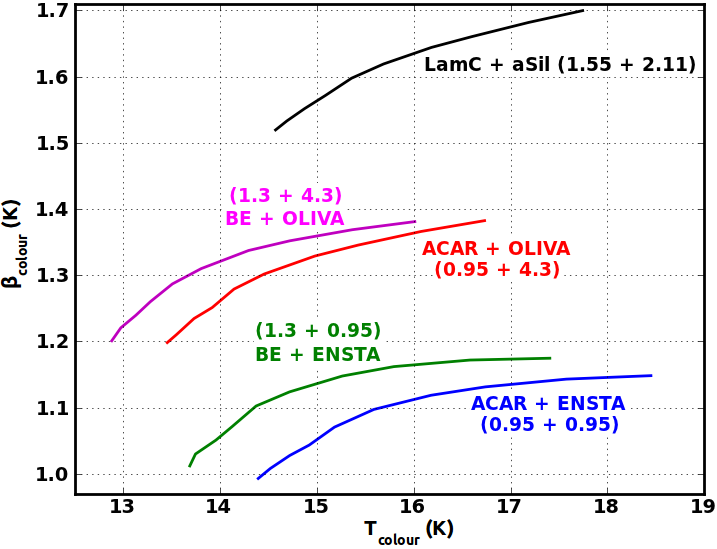}}
\caption{Emissivity spectral index as a function of colour temperature for fits of the dust SEDs between 100 and $3\,000\;\mu$m for the central pixels of the series of filaments described in Tab. 3. Line styles are the same as in Fig. \ref{beta_lambda_2}a). The numbers in parenthesis are the intrinsic opacity spectral indices of the different grain populations.}
\label{beta_T_labo} 
\end{figure}

We now consider that the ten cylindrical clouds described in Tab. 3 contain various mixtures of the ACAR, BE, OLIVA, and ENSTA grains (for more details, see also Appendix \ref{appendix_spectral_variations}). The resulting dust SEDs for these different mixtures of grain populations are shown in Fig. \ref{beta_lambda_2}a (ACAR or BE + OLIVA or ENSTA). The colour temperatures and emissivity spectral indices are determined by fitting the dust SEDs between 100 and $3\,000\;\mu$m (Fig. \ref{beta_T_labo}). The four mixtures produce lower colour temperatures than the DHGL populations, which is due to their increased emissivity in the far IR. For a central $A_V = 10$, the colour temperature is decreased by $\sim 0.2$ K for a mixture of ACAR and ENSTA, $\sim 1.2$ K for ACAR and OLIVA, $\sim 1$ K for BE and ENSTA, and $\sim 1.9$ K for BE and OLIVA. The temperature decrease is the strongest for the dust populations which have the highest absorption efficiencies in the far-IR and the lowest at longer wavelengths ($\lambda \gtrsim 1\,000-2\,000 \; \mu$m). For the dust temperature, the important spectral range is thus the far-IR, where large grains emit most of their energy in dense molecular clouds. This means that the extrapolation of the grain optical properties above 2 mm for the carbonaceous grains and $1\,500$ $\mu$m for the silicates has no influence on the estimate of the dust colour temperature. However, Fig. \ref{beta_T_labo} shows that the different values of $\beta$ in the submillimetre for OLIVA and ENSTA lead to a flatter spectral index for mixtures containing ENSTA ($1.0 \lesssim \beta \lesssim 1.2$) than for OLIVA ($1.2 \lesssim \beta \lesssim 1.4$).

For data analysis, it is useful to know whether the nature of the dust grains can be determined by observing the spectral variations in the SEDs of the clouds. To address this question, we calculate the dust emissivity using the colour temperatures estimated previously (Fig. \ref{beta_T_labo}) and the hydrogen column density $N_H$
\begin{equation}
\label{equation_emissivity}
\epsilon_{\nu} = \frac{S_{\nu}}{B_{\nu}(T_{colour}) N_H} = \frac{\tau}{N_H} \;\; {\rm in \; cm}^2{\rm /H}.
\end{equation}
Fig. \ref{beta_lambda_2} shows the example of mixtures of ACAR with either ENSTA or OLIVA grains for a cloud with $A_V = 1$. The results are highly dependent on the grain populations included in the model. In the case where the silicates are assumed to be OLIVA-like, the slope of the emissivity of the resulting SED is similar to that of the carbonaceous grains for $\lambda \gtrsim 300-1\,200 \; \mu$m depending on the cloud density, and regardless of the nature of the carbonaceous grain population (ACAR, BE, or LamC). This can be explained by the strong decrease in the emissivity of the OLIVA grains relative to that of the carbonaceous grains at long wavelengths (Fig. \ref{beta_lambda_1}). Shorter wavelengths are then too close to the peak of the SED to be used (Fig. \ref{beta_lambda_2}a). In this spectral range ($\lambda \lesssim 300 \; \mu$m), the spectral shape of the SED is strongly influenced by radiative transfer effects (i.e. by temperature mixing) and consequently no spectral variations can be associated with the grain optical properties. When the silicates are assumed to be ENSTA-like, the results are then less clear. First, for a mixture with BE grains, we cannot make any obvious correlation between the slope of the absorption efficiencies and the SED emissivity, except for $800 \lesssim \lambda \lesssim 1\,700 \; \mu$m. In this spectral range, the slope is similar to the one of ENSTA but differs at longer wavelength owing to the strong handover of the BE absorption efficiency around $2\,000\; \mu$m. Second, for a mixture with ACAR grains that have the same emissivity spectral index as ENSTA grains for $\lambda \geqslant 2\,000 \; \mu$m, the emissivity spectral index is similar to the one of ACAR for $\lambda \gtrsim 720 \; \mu$m. It is also possible to distinguish the bump of the ACAR absorption efficiency around $1\,400 \; \mu$m. We conclude that it could be possible but quite difficult to deduce the nature of the dust grain populations from the spectral variations in the dust emissivity estimated from the dust SED. As seen in Section \ref{section_noise}, the observed dust emissivity spectral index may also be affected by noise, meaning that even more confusion should be expected in real observations.

We also note that variations in the emissivity spectral index with wavelength alone can explain neither the strong decrease in the colour temperature towards dense molecular clouds nor the $\beta_{colour}-T_{colour}$ anti-correlation (see Fig. \ref{beta_T_labo}). The last one may be caused by intrinsic variations in the grain optical properties with temperature.

\subsection{Variations in optical properties with temperature: $\beta(T)$}
\label{section_beta_T}

\begin{figure}[!t]
\centerline{
\includegraphics[width=0.42\textwidth]{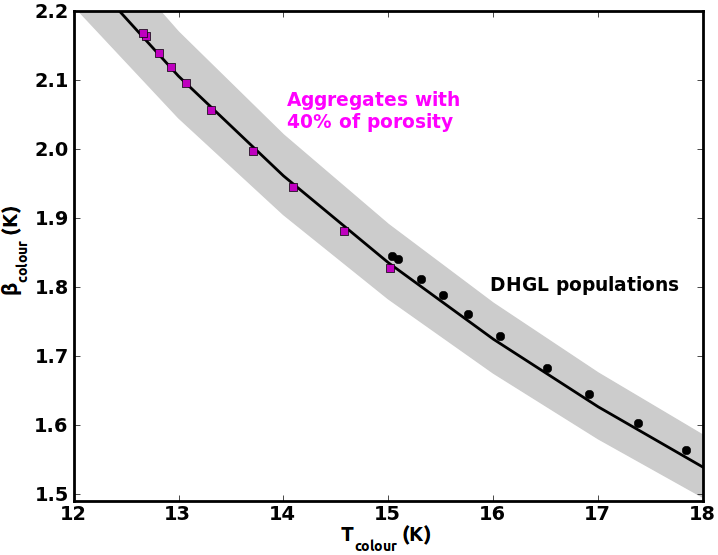}}
\caption{The solid line is the $\beta_{colour}-T_{colour}$ anti-correlation observed by \citet{PlanckMontier2011} (grey area shows the uncertainty). The black circles show the relation obtained after fitting the resulting SEDs in the Planck HFI and IRIS 100 $\mu$m bands for the central pixels of the series of filaments described in Tab. 3, when the intrinsic $\beta-T_{dust}$ relation described by Eq. \ref{fit_montier} is introduced in the $Q_{abs}$ of LamC and aSil grains. In this case, the modelled $\beta_{colour}-T_{colour}$ anti-correlation is also represented for the population of aggregates with a porosity degree of 40\% (magenta squares) and the same intrinsic relation.}
\label{beta_T_intrinsic} 
\end{figure}

To investigate the influence of intrinsic variations in the emissivity spectral index with the temperature, $\beta_{Q_{abs}}(T_{dust})$, we use the temperature-dependent grain emissivity tool of DustEM described by \citet{Compiegne2011}, which allows us to vary the spectral index of the absorption efficiency, $\beta_{Q_{abs}}$, with the grain temperature, $T_{dust}$. In DustEM, the temperature dependence is considered as a correction factor applied to a standard non-varying $Q_{abs}$
\begin{equation}
Q_{abs}(a, \nu) = Q_0(a, \nu) (\nu / \nu_t)^{\delta\beta(T_{dust}) H(\nu_t/\nu)},
\end{equation}
where $Q_0$ is the grain absorption efficiency without the temperature dependence for size $a$ and frequency $\nu$, $\nu_t = c / \lambda_t$ is the threshold frequency from which the correction applies, and $\delta\beta(T_{dust}) = \beta_{Q_{abs}}(T_{dust}) - \beta_0$ is the index correction with respect to $\beta_0$, which is the spectral index without temperature dependence (Tab. 2). The threshold function in wavelength is $H(x) = 0.5(1+4 \tanh(x)/s)$ where $x = \log(\lambda / \lambda_t)$ and $s$ controls the steepness of the transition. We choose $s = 1$ and $\lambda_t = 100 \; \mu$m and deal with the clouds described in Tab. 3 for the DHGL populations. From laboratory measurements, \citet{Mennella1998} and \citet{Coupeaud2011} demonstrated that for both amorphous carbonaceous grains and amorphous silicates the opacity decreases with decreasing temperature, and that the spectral index at a given wavelength increases when the grain temperature decreases from 300 K to 10 K. Hence we assume that for both LamC and aSil the emissivity spectral index varies with the temperature. For the sake of simplicity, we assume that the variations are similar for the two dust populations with $\beta_0 = 1.55$ for the LamC population and 2.11 for the aSil population. We do not consider the effects of noise (see Section \ref{section_noise} for a detailed description).

Most of the available observations are more or less consistent with a $\beta_{colour}-T_{colour}$ anti-correlation that can be fitted with $\beta_{colour} \propto T_{colour}^{-1}$, when considering the large error bars these fitted parameters suffer \citep{Desert2008, Paradis2010, Veneziani2010, PlanckAbergel2011, PlanckMontier2011}. For instance, \citet{PlanckMontier2011} observed starless cold clumps and found that
\begin{eqnarray}
\beta_{colour} &=& [(0.02 \pm 0.001) + (0.035 \pm 0.001) T_{colour}]^{-1} \nonumber\\
\Leftrightarrow \beta_{colour} &\sim& 1.38 \left(\frac{T_{colour}}{20\;{\rm K}}\right)^{-0.96},
\end{eqnarray}
for $7 \lesssim T_{colour} \lesssim 20$ K, when fitting the SEDs in the Planck HFI (857, 545, and 353 GHz) and IRAS 100 $\mu$m bands. To reproduce this relation, we have to introduce the intrinsic $\beta-T_{dust}$ relation in the absorption efficiency $Q_{abs}$
\begin{equation}
\label{fit_montier}
\beta_{Q_{abs}} = 1.38 \left(\frac{T_{dust}}{20\;{\rm K}}\right)^{-0.96},
\end{equation}
which allows us to mimic the observed anti-correlation for $15 \lesssim T_{colour} \lesssim 18$ K (Fig. \ref{beta_T_intrinsic}). To get lower colour temperatures, we apply the same intrinsic $\beta_{Q_{abs}}(T_{dust})$ relation to the population of aggregates with 40\% of porosity ($\beta_0 = 1.27$) yielding colour temperatures as low as 12 K and almost the same $\beta_{colour}-T_{colour}$ relation as for the DHGL populations (see Tab. 2 and Fig. \ref{beta_T_intrinsic}). We repeated this exercise in the Herschel PACS and SPIRE bands\footnote{\citet{Paradis2010} observed the Galactic plane at $l = 59\degr$ with Herschel PACS and SPIRE.} as in \citet{Paradis2010}, and in the BOOMER\tiny{AN}\normalsize{G} bands\footnote{\citet{Veneziani2010} observed cirrus at high Galactic latitude ($b\sim -40\degr$) with BOOMERANG channels at 245 and 345 GHz (Balloon Observations Of Millimetric Extragalactic Radiation and Geophysics), and in the IRAS 100 $\mu$m and DIRBE 240 $\mu$m channels (Diffuse InfraRed Background Experiment onboard COBE, COsmic Background Explorer).} as in \citet{Veneziani2010}. In the two cases, we were able to reproduce the observed anti-correlation using $\beta_{Q_{abs}}(T_{dust}) \sim \beta_{colour}(T_{colour})$, independent of the spectral bands considered.

The slope of the anti-correlation seems to be only marginally affected by radiative transfer effects for relations around $\beta_{Q_{abs}} \propto T_{dust}^{-1}$. This means that the observed relations reflect the intrinsic variations in the grain optical properties for SEDs fitted for the spectral bands of the instruments mentioned here when considering noiseless observations. This might no longer be true when the relation is shallower, $\beta_{Q_{abs}} \propto T_{dust}^{-0.6 \; {\rm to} \; 0.0}$, as the radiative transfer flattening may then win over the intrinsic anti-correlation. We also find that the steepness of the relation is almost unchanged when changing dust populations from DHGL grains to aggregates and that the change in the spectral index without temperature dependence, $\beta_0$, from one dust population to another does not produce a strong discontinuity in the observed $\beta_{colour}-T_{colour}$ relation (Fig. \ref{beta_T_intrinsic}). This indicates that the $\beta_{colour}-T_{colour}$ relations should be the same for all colour temperatures, at least as long as the clouds are starless \citep{Malinen2011}, with the highest temperatures coming from the diffuse regions and the lowest from the centre of dense molecular clouds where grain growth is expected. We do believe that our simple modelling of the intrinsic $\beta_{Q_{abs}}-T_{dust}$ anti-correlation will be a very promising ingredient of future studies including the detailed physics of the emission of amorphous solids \citep{Meny2007}, however this is beyond the scope of this paper. Finally, we note that the $\beta_{colour}-T_{colour}$ relations observed with Planck and Herschel are affected by noise, which may produce part of the anti-correlation (see Section \ref{section_noise}, Shetty et al. 2009, Blain et al. 2003). Therefore, the measured relations may overestimate the steepness of the true $\beta_{colour}-T_{colour}$ relation, where ''true'' means the relation observed after the radiative transfer effects but without the noise, and thus the steepness of the intrinsic $\beta_{Q_{abs}}(T_{dust})$ relation.

\section{Column density and dust emissivity estimates}
\label{column_density_estimates}

In addition to measurements of the dust colour temperature and emissivity spectral index, the dust SEDs from dense clouds can also be used to determine the hydrogen column density, $N_H$, or the dust emissivity at a given wavelength, $\epsilon(\lambda)$. Knowledge of the column density is important as it allows us to estimate the mass of a cloud if its distance is known. The variations in the dust emissivity from one region to another is then commonly used as a tracer of grain growth from the diffuse medium to the centre of dense molecular clouds. In this section, we examine the accuracy of the estimates of $N_H$ and $\epsilon(\lambda)$ using the dust SEDs, and also the influence of the spectral bands used to carry out this kind of analysis.

\subsection{Column density}
\label{section_column_density}

\begin{figure}[!t]
\centerline{
\includegraphics[width=0.42\textwidth]{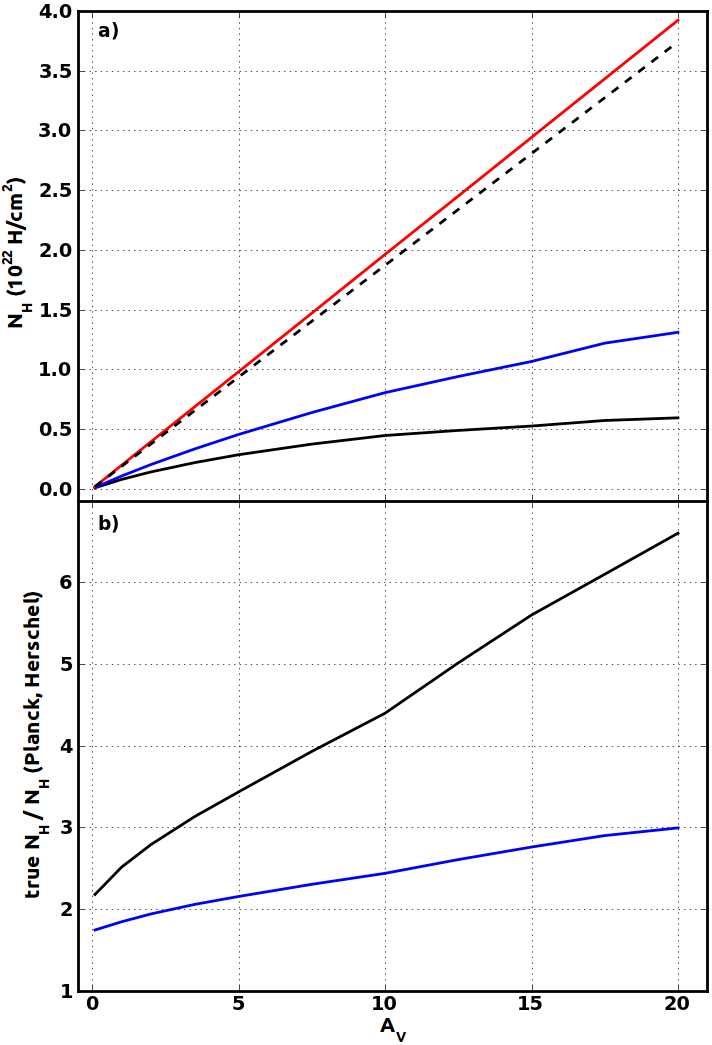}}
\caption{Case of the DHGL populations. a) True hydrogen column density $N_H$ at the centre of the modelled clouds as a function of the central visual extinction $A_V$ (red line). The dashed line shows the $N_H-A_V$ relation measured by \citet{Bohlin1978} for the diffuse ISM. The blue line represents the column density estimated from the SED measured in the Planck-HFI and IRIS 100 $\mu$m bands, whereas the black line comes from the SED-fits made in the Herschel bands. b) Ratio of the true column density of the modelled clouds to the column densities measured in Herschel and Planck bands (black and blue, respectively).}
\label{column_density} 
\end{figure}

Using the colour temperatures determined with the fits of the dust SEDs, the column densities of the clouds can be expressed as
\begin{equation}
\label{equation_NH}
N_H = \frac{S_{\nu}}{B_{\nu}(T_{colour}) \, \kappa_{\nu} \, \mu m_H},
\end{equation}
where $\mu = 1.33$ is the mean atomic mass, $m_H$ is the mass of the hydrogen atom, $S_{\nu}$ is the emerging intensity at frequency $\nu$, $B_{\nu}(T)$ is the Planck function at temperature $T$ and frequency $\nu$, and $\kappa_{\nu} = \epsilon / \mu m_H$ is the dust opacity at frequency $\nu$. We calculate the observed column densities for SEDs fitted in the Herschel PACS and SPIRE bands, and then in the Planck HFI and IRIS 100 $\mu$m bands (see Section \ref{effect_of_radiative_transfer}). For the Planck estimate of $N_H$, we use the band at 857 GHz, and for the Herschel counterpart the band at 250 $\mu$m. The results for the ten clouds described in Tab. 3, containing DHGL grain populations, are shown in Fig. \ref{column_density}. In this case, the opacity at 250 $\mu$m is equal to 0.051 cm$^2$/g and at 857 GHz to 0.025 cm$^2$/g (Tab. 2). The column density is strongly underestimated with both Planck and Herschel bands, and this discrepancy increases with increasing $A_V = 1$ to 20 from a factor of 1.8 to 3.0 in the Planck case, and from 2.5 to 6.6 in the Herschel case. This is in good agreement with previous studies \citep{Evans2001, Stamatellos2003, Fischera2011, Malinen2011}. This difference comes from the assumption of a single temperature in Eq. \ref{equation_NH}. The colour temperature is always higher than the temperature of the grains at the centre of the clouds and this overestimate increases with the cloud central $A_V$ as can be seen in Fig. \ref{Tcolour_Tdust}. The influence of the difference between the colour temperature and the temperature of the grains on the calculation of $N_H$ comes from the Planck function: for instance, at 857 GHz we found that $B(18 \;{\rm K})/B(14 \;{\rm K}) \sim 2$, and at 250 $\mu$m that $B(18 \;{\rm K})/B(14 \;{\rm K}) \sim 2.6$, which explains why the estimate of $N_H$ with the Planck 857 GHz band is slightly more accurate than with the Herschel 250 $\mu$m band. These ratios increase to 9.5 and 22, respectively, when the temperature differs from 8 K to 14 K. For comparison with more diffuse media and to test the influence of the radiative transfer, we modelled a cloud with a central density equal to 30 H/cm$^3$, leading to a central visual extinction $A_V = 0.09$. The column density is underestimated by a factor of 1.7 for Planck and 2.2 for Herschel. In this case, the discrepancy is not explained by radiative transfer effects but only by the mixture of several dust populations distributed in size and having different temperatures.

Linear fits of the $N_H-A_V$ relations corresponding to the different populations of grains used in this paper, and for the cloud geometry and density distributions described in Section \ref{section_cloud_geometry}, are listed in Tab. 5. This is given for comparison with \citet{Bohlin1978} ($N_H = 1.87 \times 10^{21} A_V$ H/cm$^2$). Finally, real estimates of the column density from far IR and submillimetre observations are usually worsened by the ignorance of the true dust opacity inside dense molecular clouds.

\begin{table}
\label{NH_Av}
\centering
\caption{Relation between the hydrogen column density and the visual extinction for the dust populations listed in Tab. 2.}
\begin{tabular}{lc}
\hline
\hline
Dust populations & Linear fits for the $N_H-A_V$ relation \\
\hline
DHGL             & $N_H \sim 1.97 \times 10^{21} A_V$ \\
Accreted grains  & $N_H \sim 1.80 \times 10^{21} A_V$ \\
Aggregates 0\%   & $N_H \sim 2.13 \times 10^{21} A_V$ \\
Aggregates 10\%  & $N_H \sim 2.05 \times 10^{21} A_V$ \\
Aggregates 25\%  & $N_H \sim 1.90 \times 10^{21} A_V$ \\
Aggregates 40\%  & $N_H \sim 1.76 \times 10^{21} A_V$ \\
\hline
\end{tabular}
\end{table}

\subsection{Dust emissivity}
\label{section_dust_emissivity}

\begin{figure}[!t]
\centerline{
\includegraphics[width=0.42\textwidth]{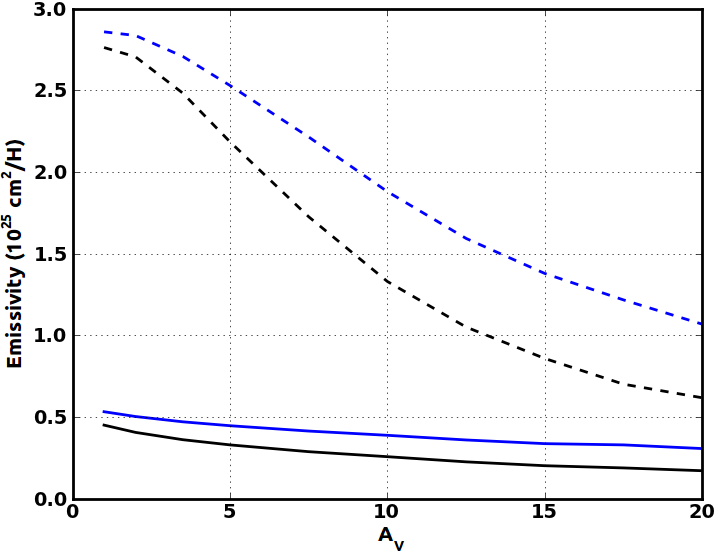}}
\caption{Case of the DHGL grains (solid lines) and of the aggregates with 25\% of porosity (dashed lines). Dust emissivity at 250 $\mu$m as a function of the central visual extinction $A_V$. The black and the blue lines show the emissivity measured in the Herschel and in the Planck-HFI bands respectively for the central pixels of the filaments described in Tab. 3, assuming the observer does know the true column densities of the clouds.}
\label{emissivity} 
\end{figure}

As we know the true column densities of the modelled clouds, we can measure the dust emissivity using Eq. \ref{equation_emissivity}. As in the previous section, using the ten clouds in Tab. 3, we perform this analysis with the SED fitting in both Planck and Herschel bands. Figure \ref{emissivity} presents the variations in the dust emissivity at 250 $\mu$m as a function of the visual extinction for the DHGL populations and for aggregates with a porosity degree of 40\%. In an opposite way to the variations observed in the ISM \citep{Juvela2011, PlanckAbergel2011}, the emissivity decreases when the cloud density increases. This is due to the overestimate of the dust temperature made when using the colour temperature, which was already seen in Section \ref{section_column_density} for the measurement of $N_H$. The use of the grain temperatures at the centre of the clouds, as discussed in Section \ref{section_grain_temperature}, would lead to an almost flat relation. This shows that the increase in the submillimetre emissivity towards dense clouds can be explained by neither purely radiative transfer effects nor the bias introduced by the use of the colour temperature. It must be caused by variations in the grain properties from the diffuse to the dense medium. Grain growth may explain the higher emissivity towards dense regions, as shown in Fig. \ref{emissivity}, in terms of the variations in the emissivity of aggregates with 25\% of porosity. For $A_V = 10$, the emissivity is increased by a factor of 4.8 compared with that of the DHGL grains. This factor is comparable to that observed in dense clouds with similar central $A_V$ (see Tab. 1, Juvela et al. 2011, Stepnik et al. 2003).

\section{Conclusion}
\label{conclusion}

The primary aims of this paper were to differentiate between the effects of the radiative transfer and dust properties in dense molecular clouds. Secondly we have explored some explanations of the dust emission features observed towards these regions. To do so, we modelled the dust emission from idealized filaments, which were described as infinite circular cylinders at equilibrium that were externally heated by the isotropic ISRF, including full radiative transfer calculations. We also briefly considered the case of Bonnor-Ebert spheres (Appendix \ref{appendix_BE}). Our main results are as follows:
\begin{enumerate}
\item Radiative transfer effects, i.e. the mixing of different temperatures along the line-of-sight, can explain neither the low colour temperatures measured at the centre of dense and starless molecular clouds, nor the observed $\beta_{colour}-T_{colour}$ anti-correlation. The effect of radiative transfer is to produce an artificial correlation (not an anti-correlation as observed) between the emissivity spectral index, $\beta_{colour}$, and the colour temperature.
\item For externally heated clouds with simple geometries (cylinders and spheres) containing dust populations with standard optical properties, the $\beta_{colour}-T_{colour}$ relations observed with Planck and Herschel are too steep to be explained solely by noise. For starless interstellar clouds, it is thus likely that the observed anti-correlation is related to intrinsic variations in the grain optical properties with temperature as suggested by laboratory experiments \citep{Mennella1998, Coupeaud2011} and models of the emission of amorphous solids \citep{Meny2007}. However, an anti-correlation might also be produced if the sample includes map pixels with strong internal sources or (even a small number of) clouds observed with very low S/N's.
\item The increase in the submillimetre emissivity from the diffuse to the dense medium cannot be explained by radiative transfer effects, but must originate in variations in the grain optical properties, which could be caused by their growth inside dense molecular clouds to form porous aggregates \citep{Ossenkopf1994}.
\item Regarding grain growth, only the aggregation of grains in mixed aggregates of small and large grains can explain the deficit in emission in the mid-IR, the low colour temperature, and the higher submillimetre emissivity observed towards dense molecular clouds.
\item We found that for clouds with a central visual extinction less than 10 magnitudes, or central densities less than $2 \times 10^5$ H/cm$^3$ in the case of DHGL populations, a linear relation between the colour temperature and the "true" dust temperature at the centre of the clouds does exist. Assuming that a reliable measure of the cloud central visual extinction/density is available, it is possible to estimate the dust temperature using this relation.
\item We investigated the impact of intrinsic variations in the grain optical properties with wavelength, $\beta(\lambda)$. Even if possible, we showed that it should be rather difficult to track back the nature of the dust grains from the spectral variations in the dust SEDs emerging from dense clouds. This difficulty is due mostly to the radiative transfer effects for $\lambda \lesssim 300 \; \mu$m and the mixture of different grain populations at longer wavelengths. In real observations, the situation is further confused by the presence of noise.
\item We also investigated the impact of intrinsic variations in the grain optical properties with temperature, $\beta(T)$. For the three observational results considered in this paper, with $\beta_{colour} \sim T_{colour}^{-1}$, the observed anti-correlation seems to reflect the $\beta-T_{dust}$ intrinsic relation. This result may no longer hold for shallower relations for which the radiative transfer flattening effect would win over the intrinsic anti-correlation. As noise may produce an anti-correlation, the observed relations may overestimate the steepness of the true $\beta_{colour}-T_{colour}$ relations.
\item Finally, we confirmed that the column density is strongly underestimated when determined by SED fitting. In diffuse media, it is reduced by a factor of 1.7 and 2.2 in Planck and Herschel bands, respectively. This is explained by the mixture of several dust populations widely distributed in both size and temperature. This discrepancy then increases when the central visual extinction of the clouds increases because of the difference between the colour temperature and the "true" dust temperature in the central layers of the clouds, where most of the material lies.
\end{enumerate}

\acknowledgements{The authors thank the anonymous referee for useful comments that helped us to improve the content of the paper. We also thank V. Mennella and A. Coupeaud for making their laboratory data available. N.Y., M.J., and J.M. acknowledge support from Academy of Finland projects 127015 and 1206049. J.M. also acknowledges a grant from Magnus Ehrnrooth foundation.}

\bibliography{biblio}

\begin{thebibliography}{84}
\expandafter\ifx\csname natexlab\endcsname\relax\def\natexlab#1{#1}\fi

\bibitem[{{Abergel} {et~al.}(2010){Abergel}, {Arab}, {Compi{\`e}gne}, {Kirk},
  {Ade}, {Anderson}, {Andr{\'e}}, {Baluteau}, {Bernard}, {Blagrave},
  {Bontemps}, {Boulanger}, {Cohen}, {Cox}, {Dartois}, {Davis}, {Emery},
  {Fulton}, {Gry}, {Habart}, {Huang}, {Joblin}, {Jones}, {Lagache}, {Lim},
  {Madden}, {Makiwa}, {Martin}, {Miville-Desch{\^e}nes}, {Molinari}, {Moseley},
  {Motte}, {Naylor}, {Okumura}, {Pinheiro Gon{\c c}alves}, {Polehampton},
  {Rodon}, {Russeil}, {Saraceno}, {Sauvage}, {Sidher}, {Spencer}, {Swinyard},
  {Ward-Thompson}, {White}, \& {Zavagno}}]{Abergel2010}
{Abergel}, A., {Arab}, H., {Compi{\`e}gne}, M., {et~al.} 2010, \aap, 518, L96+

\bibitem[{{Agladze} {et~al.}(1996){Agladze}, {Sievers}, {Jones}, {Burlitch}, \&
  {Beckwith}}]{Agladze1996}
{Agladze}, N.~I., {Sievers}, A.~J., {Jones}, S.~A., {Burlitch}, J.~M., \&
  {Beckwith}, S.~V.~W. 1996, \apj, 462, 1026

\bibitem[{{Alves} {et~al.}(1998){Alves}, {Lada}, {Lada}, {Kenyon}, \&
  {Phelps}}]{Alves1998}
{Alves}, J., {Lada}, C.~J., {Lada}, E.~A., {Kenyon}, S.~J., \& {Phelps}, R.
  1998, \apj, 506, 292

\bibitem[{{Andr{\'e}} {et~al.}(2010){Andr{\'e}}, {Men'shchikov}, {Bontemps},
  {K{\"o}nyves}, {Motte}, {Schneider}, {Didelon}, {Minier}, {Saraceno},
  {Ward-Thompson}, {di Francesco}, {White}, {Molinari}, {Testi}, {Abergel},
  {Griffin}, {Henning}, {Royer}, {Mer{\'{\i}}n}, {Vavrek}, {Attard},
  {Arzoumanian}, {Wilson}, {Ade}, {Aussel}, {Baluteau}, {Benedettini},
  {Bernard}, {Blommaert}, {Cambr{\'e}sy}, {Cox}, {di Giorgio}, {Hargrave},
  {Hennemann}, {Huang}, {Kirk}, {Krause}, {Launhardt}, {Leeks}, {Le Pennec},
  {Li}, {Martin}, {Maury}, {Olofsson}, {Omont}, {Peretto}, {Pezzuto}, {Prusti},
  {Roussel}, {Russeil}, {Sauvage}, {Sibthorpe}, {Sicilia-Aguilar}, {Spinoglio},
  {Waelkens}, {Woodcraft}, \& {Zavagno}}]{Andre2010}
{Andr{\'e}}, P., {Men'shchikov}, A., {Bontemps}, S., {et~al.} 2010, \aap, 518,
  L102+

\bibitem[{{Arzoumanian} {et~al.}(2011){Arzoumanian}, {Andr{\'e}}, {Didelon},
  {Konyves}, {Schneider}, {Men'shchikov}, {Sousbie}, {Zavagno}, {Bontemps}, {Di
  Francesco}, {Griffin}, {Hennemann}, {Hill1}, {Kirk}, {Martin}, {Minier},
  {Molinari}, {Motte}, {Peretto}, {Pezzuto}, {Spinoglio}, {Ward-Thompson},
  {White}, \& {Wilson}}]{Arzoumanian2011}
{Arzoumanian}, D., {Andr{\'e}}, P., {Didelon}, P., {et~al.} 2011, ArXiv
  e-prints

\bibitem[{{Bernard} {et~al.}(1999){Bernard}, {Abergel}, {Ristorcelli}, {Pajot},
  {Torre}, {Boulanger}, {Giard}, {Lagache}, {Serra}, {Lamarre}, {Puget},
  {Lepeintre}, \& {Cambr{\'e}sy}}]{Bernard1999}
{Bernard}, J.~P., {Abergel}, A., {Ristorcelli}, I., {et~al.} 1999, \aap, 347,
  640

\bibitem[{{Bernard} {et~al.}(1992){Bernard}, {Boulanger}, {Desert}, \&
  {Puget}}]{Bernard1992}
{Bernard}, J.~P., {Boulanger}, F., {Desert}, F.~X., \& {Puget}, J.~L. 1992,
  \aap, 263, 258

\bibitem[{{Blain} {et~al.}(2003){Blain}, {Barnard}, \& {Chapman}}]{Blain2003}
{Blain}, A.~W., {Barnard}, V.~E., \& {Chapman}, S.~C. 2003, \mnras, 338, 733

\bibitem[{{Bohlin} {et~al.}(1978){Bohlin}, {Savage}, \& {Drake}}]{Bohlin1978}
{Bohlin}, R.~C., {Savage}, B.~D., \& {Drake}, J.~F. 1978, \apj, 224, 132

\bibitem[{Bohren \& Huffman(1983)}]{Bohren1983}
Bohren, C.~F. \& Huffman, D.~R. 1983, New York: Wiley

\bibitem[{{Bonnor}(1956)}]{Bonnor1956}
{Bonnor}, W.~B. 1956, \mnras, 116, 351

\bibitem[{{Boudet} {et~al.}(2005){Boudet}, {Mutschke}, {Nayral}, {J{\"a}ger},
  {Bernard}, {Henning}, \& {Meny}}]{Boudet2005}
{Boudet}, N., {Mutschke}, H., {Nayral}, C., {et~al.} 2005, \apj, 633, 272

\bibitem[{{Bracco} {et~al.}(2011){Bracco}, {Cooray}, {Veneziani}, {Amblard},
  {Serra}, {Wardlow}, {Thompson}, {White}, {Auld}, {Baes}, {Bertoldi},
  {Buttiglione}, {Cava}, {Clements}, {Dariush}, {de Zotti}, {Dunne}, {Dye},
  {Eales}, {Fritz}, {Gomez}, {Hopwood}, {Ibar}, {Ivison}, {Jarvis}, {Lagache},
  {Lee}, {Leeuw}, {Maddox}, {Micha{\l}owski}, {Pearson}, {Pohlen}, {Rigby},
  {Rodighiero}, {Smith}, {Temi}, {Vaccari}, \& {van der Werf}}]{Bracco2011}
{Bracco}, A., {Cooray}, A., {Veneziani}, M., {et~al.} 2011, \mnras, 412, 1151

\bibitem[{{Cambr{\'e}sy} {et~al.}(2001){Cambr{\'e}sy}, {Boulanger}, {Lagache},
  \& {Stepnik}}]{Cambresy2001}
{Cambr{\'e}sy}, L., {Boulanger}, F., {Lagache}, G., \& {Stepnik}, B. 2001,
  \aap, 375, 999

\bibitem[{{Campeggio} {et~al.}(2007){Campeggio}, {Strafella}, {Maiolo}, {Elia},
  \& {Aiello}}]{Campeggio2007}
{Campeggio}, L., {Strafella}, F., {Maiolo}, B., {Elia}, D., \& {Aiello}, S.
  2007, \apj, 668, 316

\bibitem[{{Cardelli} \& {Clayton}(1991)}]{Cardelli1991}
{Cardelli}, J.~A. \& {Clayton}, G.~C. 1991, \aj, 101, 1021

\bibitem[{{Colangeli} {et~al.}(1995){Colangeli}, {Mennella}, {Palumbo},
  {Rotundi}, \& {Bussoletti}}]{Colangeli1995}
{Colangeli}, L., {Mennella}, V., {Palumbo}, P., {Rotundi}, A., \& {Bussoletti},
  E. 1995, \aaps, 113, 561

\bibitem[{{Compi{\`e}gne} {et~al.}(2011){Compi{\`e}gne}, {Verstraete}, {Jones},
  {Bernard}, {Boulanger}, {Flagey}, {Le Bourlot}, {Paradis}, \&
  {Ysard}}]{Compiegne2011}
{Compi{\`e}gne}, M., {Verstraete}, L., {Jones}, A., {et~al.} 2011, \aap, 525,
  A103+

\bibitem[{{Coupeaud} {et~al.}(2011){Coupeaud}, {Demyk}, {Meny}, {Nayral},
  {Delpech}, {Leroux}, {Depecker}, {Creff}, {Brubach}, \& {Roy}}]{Coupeaud2011}
{Coupeaud}, A., {Demyk}, K., {Meny}, C., {et~al.} 2011,
  ArXiv:astro-ph/1109.2758

\bibitem[{{Dapp} \& {Basu}(2009)}]{Dapp2009}
{Dapp}, W.~B. \& {Basu}, S. 2009, \mnras, 395, 1092

\bibitem[{{D{\'e}sert} {et~al.}(2008){D{\'e}sert}, {Mac{\'{\i}}as-P{\'e}rez},
  {Mayet}, {Giardino}, {Renault}, {Aumont}, {Beno{\^i}t}, {Bernard},
  {Ponthieu}, \& {Tristram}}]{Desert2008}
{D{\'e}sert}, F.-X., {Mac{\'{\i}}as-P{\'e}rez}, J.~F., {Mayet}, F., {et~al.}
  2008, \aap, 481, 411

\bibitem[{{Dorschner} \& {Henning}(1995)}]{Dorschner1995}
{Dorschner}, J. \& {Henning}, T. 1995, \aapr, 6, 271

\bibitem[{{Draine}(2003{\natexlab{a}})}]{DD2003}
{Draine}, B.~T. 2003{\natexlab{a}}, \araa, 41, 241

\bibitem[{{Draine}(2003{\natexlab{b}})}]{Draine2003}
{Draine}, B.~T. 2003{\natexlab{b}}, \apj, 598, 1026

\bibitem[{{Draine} \& {Lee}(1984)}]{Draine1984}
{Draine}, B.~T. \& {Lee}, H.~M. 1984, \apj, 285, 89

\bibitem[{{Dupac} {et~al.}(2003){Dupac}, {Bernard}, {Boudet}, {Giard},
  {Lamarre}, {M{\'e}ny}, {Pajot}, {Ristorcelli}, {Serra}, {Stepnik}, \&
  {Torre}}]{Dupac2003}
{Dupac}, X., {Bernard}, J.-P., {Boudet}, N., {et~al.} 2003, \aap, 404, L11

\bibitem[{{Ebert}(1955)}]{Ebert1955}
{Ebert}, R. 1955, \zap, 37, 217

\bibitem[{{Evans} {et~al.}(2001){Evans}, {Rawlings}, {Shirley}, \&
  {Mundy}}]{Evans2001}
{Evans}, II, N.~J., {Rawlings}, J.~M.~C., {Shirley}, Y.~L., \& {Mundy}, L.~G.
  2001, \apj, 557, 193

\bibitem[{{Fiege} \& {Pudritz}(2000)}]{Fiege2000}
{Fiege}, J.~D. \& {Pudritz}, R.~E. 2000, \mnras, 311, 85

\bibitem[{{Fischera}(2011)}]{Fischera2011}
{Fischera}, J. 2011, \aap, 526, A33+

\bibitem[{{Fischera} \& {Dopita}(2008)}]{Fischera2008}
{Fischera}, J. \& {Dopita}, M.~A. 2008, \apjs, 176, 164

\bibitem[{{Fitzpatrick} \& {Massa}(1988)}]{Fitzpatrick1988}
{Fitzpatrick}, E.~L. \& {Massa}, D. 1988, \apj, 328, 734

\bibitem[{{Fitzpatrick} \& {Massa}(2007)}]{Fitzpatrick2007}
{Fitzpatrick}, E.~L. \& {Massa}, D. 2007, \apj, 663, 320

\bibitem[{{Flagey} {et~al.}(2009){Flagey}, {Noriega-Crespo}, {Boulanger},
  {Carey}, {Brooke}, {Falgarone}, {Huard}, {McCabe}, {Miville-Desch{\^e}nes},
  {Padgett}, {Paladini}, \& {Rebull}}]{Flagey2009}
{Flagey}, N., {Noriega-Crespo}, A., {Boulanger}, F., {et~al.} 2009, \apj, 701,
  1450

\bibitem[{Fogel \& Leung(1998)}]{Fogel1998}
Fogel, M.~E. \& Leung, C.~M. 1998, \apj, 501, 175

\bibitem[{{Hill} {et~al.}(2011){Hill}, {Motte}, {Didelon}, {Bontemps},
  {Minier}, {Hennemann}, {Schneider}, {Andr{\'e}}, {Men`Shchikov}, {Anderson},
  {Arzoumanian}, {Bernard}, {di Francesco}, {Elia}, {Giannini}, {Griffin},
  {K{\"o}nyves}, {Kirk}, {Marston}, {Martin}, {Molinari}, {Nguyn Lu'O'Ng},
  {Peretto}, {Pezzuto}, {Roussel}, {Sauvage}, {Sousbie}, {Testi},
  {Ward-Thompson}, {White}, {Wilson}, \& {Zavagno}}]{Hill2011}
{Hill}, T., {Motte}, F., {Didelon}, P., {et~al.} 2011, \aap, 533, A94+

\bibitem[{{Juvela}(2005)}]{Juvela2005}
{Juvela}, M. 2005, \aap, 440, 531

\bibitem[{{Juvela} \& {Padoan}(2003)}]{Juvela2003}
{Juvela}, M. \& {Padoan}, P. 2003, \aap, 397, 201

\bibitem[{{Juvela} {et~al.}(2010){Juvela}, {Ristorcelli}, {Montier},
  {Marshall}, {Pelkonen}, {Malinen}, {Ysard}, {T{\'o}th}, {Harju}, {Bernard},
  {Schneider}, {Vereb{\'e}lyi}, {Anderson}, {Andr{\'e}}, {Giard}, {Krause},
  {Lehtinen}, {Macias-Perez}, {Martin}, {McGehee}, {Meny}, {Motte}, {Pagani},
  {Paladini}, {Reach}, {Valenziano}, {Ward-Thompson}, \&
  {Zavagno}}]{Juvela2010}
{Juvela}, M., {Ristorcelli}, I., {Montier}, L.~A., {et~al.} 2010, \aap, 518,
  L93+

\bibitem[{{Juvela} {et~al.}(2012){Juvela}, {Ristorcelli}, {Pagani}, {Doi},
  {Pelkonen}, {Marshall}, {Bernard}, {Falgarone}, {Malinen}, {Marton},
  {McGehee}, {Montier}, {Motte}, {Paladini}, {Toth}, {Ysard}, {Zahorecz}, \&
  {Zavagno}}]{Juvela2012b}
{Juvela}, M., {Ristorcelli}, I., {Pagani}, L., {et~al.} 2012, ArXiv 1202.1672

\bibitem[{{Juvela} {et~al.}(2011){Juvela}, {Ristorcelli}, {Pelkonen},
  {Marshall}, {Montier}, {Bernard}, {Paladini}, {Lunttila}, {Abergel},
  {Andr{\'e}}, {Dickinson}, {Dupac}, {Malinen}, {Martin}, {McGehee}, {Pagani},
  {Ysard}, \& {Zavagno}}]{Juvela2011}
{Juvela}, M., {Ristorcelli}, I., {Pelkonen}, V.-M., {et~al.} 2011, \aap, 527,
  A111+

\bibitem[{{Juvela} \& {Ysard}(2012)}]{Juvela2012}
{Juvela}, M. \& {Ysard}, N. 2012, \aap, 539, A71

\bibitem[{{Kemper} {et~al.}(2004){Kemper}, {Vriend}, \& {Tielens}}]{Kemper2004}
{Kemper}, F., {Vriend}, W.~J., \& {Tielens}, A.~G.~G.~M. 2004, \apj, 609, 826

\bibitem[{{Kim} {et~al.}(1994){Kim}, {Martin}, \& {Hendry}}]{Kim1994}
{Kim}, S., {Martin}, P.~G., \& {Hendry}, P.~D. 1994, \apj, 422, 164

\bibitem[{{Kiss} {et~al.}(2006){Kiss}, {{\'A}brah{\'a}m}, {Laureijs},
  {Mo{\'o}r}, \& {Birkmann}}]{Kiss2006}
{Kiss}, C., {{\'A}brah{\'a}m}, P., {Laureijs}, R.~J., {Mo{\'o}r}, A., \&
  {Birkmann}, S.~M. 2006, \mnras, 373, 1213

\bibitem[{K{\"o}hler {et~al.}(2011)K{\"o}hler, Guillet, \& Jones}]{Kohler2011}
K{\"o}hler, M., Guillet, V., \& Jones, A. 2011, Astronomy {\&} Astrophysics,
  528, 96

\bibitem[{{Kramer} {et~al.}(2003){Kramer}, {Richer}, {Mookerjea}, {Alves}, \&
  {Lada}}]{Kramer2003}
{Kramer}, C., {Richer}, J., {Mookerjea}, B., {Alves}, J., \& {Lada}, C. 2003,
  \aap, 399, 1073

\bibitem[{{Lehtinen} {et~al.}(2007){Lehtinen}, {Juvela}, {Mattila}, {Lemke}, \&
  {Russeil}}]{Lehtinen2007}
{Lehtinen}, K., {Juvela}, M., {Mattila}, K., {Lemke}, D., \& {Russeil}, D.
  2007, \aap, 466, 969

\bibitem[{Mackowski(2006)}]{Mackowski2006}
Mackowski, D.~W. 2006, Journal of Quantitative Spectroscopy and Radiative
  Transfer, 100, 237

\bibitem[{{Malinen} {et~al.}(2011){Malinen}, {Juvela}, {Collins}, {Lunttila},
  \& {Padoan}}]{Malinen2011}
{Malinen}, J., {Juvela}, M., {Collins}, D.~C., {Lunttila}, T., \& {Padoan}, P.
  2011, \aap, 530, A101+

\bibitem[{{Mathis} {et~al.}(1983){Mathis}, {Mezger}, \& {Panagia}}]{Mathis1983}
{Mathis}, J.~S., {Mezger}, P.~G., \& {Panagia}, N. 1983, \aap, 128, 212

\bibitem[{Mathis \& Whiffen(1989)}]{Mathis1989}
Mathis, J.~S. \& Whiffen, G. 1989, \apj, 341, 808

\bibitem[{{Mennella} {et~al.}(1998){Mennella}, {Brucato}, {Colangeli},
  {Palumbo}, {Rotundi}, \& {Bussoletti}}]{Mennella1998}
{Mennella}, V., {Brucato}, J.~R., {Colangeli}, L., {et~al.} 1998, \apj, 496,
  1058

\bibitem[{{Meny} {et~al.}(2007){Meny}, {Gromov}, {Boudet}, {Bernard},
  {Paradis}, \& {Nayral}}]{Meny2007}
{Meny}, C., {Gromov}, V., {Boudet}, N., {et~al.} 2007, \aap, 468, 171

\bibitem[{{Miville-Desch{\^e}nes} {et~al.}(2010){Miville-Desch{\^e}nes},
  {Martin}, {Abergel}, {Bernard}, {Boulanger}, {Lagache}, {Anderson},
  {Andr{\'e}}, {Arab}, {Baluteau}, {Blagrave}, {Bontemps}, {Cohen},
  {Compiegne}, {Cox}, {Dartois}, {Davis}, {Emery}, {Fulton}, {Gry}, {Habart},
  {Huang}, {Joblin}, {Jones}, {Kirk}, {Lim}, {Madden}, {Makiwa}, {Menshchikov},
  {Molinari}, {Moseley}, {Motte}, {Naylor}, {Okumura}, {Pinheiro Gon{\c
  c}alves}, {Polehampton}, {Rod{\'o}n}, {Russeil}, {Saraceno}, {Schneider},
  {Sidher}, {Spencer}, {Swinyard}, {Ward-Thompson}, {White}, \&
  {Zavagno}}]{MAMD2010}
{Miville-Desch{\^e}nes}, M., {Martin}, P.~G., {Abergel}, A., {et~al.} 2010,
  \aap, 518, L104+

\bibitem[{{Miville-Desch{\^e}nes} \& {Lagache}(2005)}]{Miville2005}
{Miville-Desch{\^e}nes}, M.-A. \& {Lagache}, G. 2005, \apjs, 157, 302

\bibitem[{{Nakamura} \& {Umemura}(1999)}]{Nakamura1999}
{Nakamura}, F. \& {Umemura}, M. 1999, \apj, 515, 239

\bibitem[{{Nguyen Luong} {et~al.}(2011){Nguyen Luong}, {Motte}, {Hennemann},
  {Hill}, {Rygl}, {Schneider}, {Bontemps}, {Men'shchikov}, {Andr{\'e}},
  {Peretto}, {Anderson}, {Arzoumanian}, {Deharveng}, {Didelon}, {Di Francesco},
  {Griffin}, {Kirk}, {Konyves}, {Martin}, {Maury}, {Minier}, {Molinari},
  {Pestalozzi}, {Pezzuto}, {Reid}, {Roussel}, {Schuller}, {Testi},
  {Ward-Thompson}, {White}, \& {Zavagno}}]{Nguyen2011}
{Nguyen Luong}, Q., {Motte}, F., {Hennemann}, M., {et~al.} 2011, ArXiv e-prints

\bibitem[{{Ossenkopf}(1993)}]{Ossenkopf1993}
{Ossenkopf}, V. 1993, \aap, 280, 617

\bibitem[{{Ossenkopf} \& {Henning}(1994)}]{Ossenkopf1994}
{Ossenkopf}, V. \& {Henning}, T. 1994, \aap, 291, 943

\bibitem[{{Ostriker}(1964)}]{Ostriker1964}
{Ostriker}, J. 1964, \apj, 140, 1056

\bibitem[{{Paradis} {et~al.}(2009){Paradis}, {Bernard}, \&
  {M{\'e}ny}}]{Paradis2009}
{Paradis}, D., {Bernard}, J.-P., \& {M{\'e}ny}, C. 2009, \aap, 506, 745

\bibitem[{{Paradis} {et~al.}(2011){Paradis}, {Bernard}, {M{\'e}ny}, \&
  {Gromov}}]{Paradis2011}
{Paradis}, D., {Bernard}, J.-P., {M{\'e}ny}, C., \& {Gromov}, V. 2011, \aap,
  534, A118

\bibitem[{{Paradis} {et~al.}(2010){Paradis}, {Veneziani}, {Noriega-Crespo},
  {Paladini}, {Piacentini}, {Bernard}, {de Bernardis}, {Calzoletti},
  {Faustini}, {Martin}, {Masi}, {Montier}, {Natoli}, {Ristorcelli}, {Thompson},
  {Traficante}, \& {Molinari}}]{Paradis2010}
{Paradis}, D., {Veneziani}, M., {Noriega-Crespo}, A., {et~al.} 2010, \aap, 520,
  L8+

\bibitem[{{Planck Collaboration}(2011{\natexlab{a}})}]{PlanckBernard2011}
{Planck Collaboration}. 2011{\natexlab{a}}, \aap, 536, A19

\bibitem[{{Planck Collaboration}(2011{\natexlab{b}})}]{PlanckRistorcelli2011}
{Planck Collaboration}. 2011{\natexlab{b}}, \aap, 536, A22

\bibitem[{{Planck Collaboration}(2011{\natexlab{c}})}]{PlanckMontier2011}
{Planck Collaboration}. 2011{\natexlab{c}}, \aap, 536, A23

\bibitem[{{Planck Collaboration}(2011{\natexlab{d}})}]{PlanckAbergel2011}
{Planck Collaboration}. 2011{\natexlab{d}}, \aap, 536, A25

\bibitem[{{Ridderstad} \& {Juvela}(2010)}]{Ridderstad2010}
{Ridderstad}, M. \& {Juvela}, M. 2010, \aap, 520, A18+

\bibitem[{{Ridderstad} {et~al.}(2006){Ridderstad}, {Juvela}, {Lehtinen},
  {Lemke}, \& {Liljestr{\"o}m}}]{Ridderstad2006}
{Ridderstad}, M., {Juvela}, M., {Lehtinen}, K., {Lemke}, D., \&
  {Liljestr{\"o}m}, T. 2006, \aap, 451, 961

\bibitem[{{Rod{\'o}n} {et~al.}(2010){Rod{\'o}n}, {Zavagno}, {Baluteau},
  {Anderson}, {Polehampton}, {Abergel}, {Motte}, {Bontemps}, {Ade},
  {Andr{\'e}}, {Arab}, {Beichman}, {Bernard}, {Blagrave}, {Boulanger}, {Cohen},
  {Compiegne}, {Cox}, {Dartois}, {Davis}, {Emery}, {Fulton}, {Gry}, {Habart},
  {Halpern}, {Huang}, {Joblin}, {Jones}, {Kirk}, {Lagache}, {Lin}, {Madden},
  {Makiwa}, {Martin}, {Miville-Desch{\^e}nes}, {Molinari}, {Moseley}, {Naylor},
  {Okumura}, {Orieux}, {Pinheiro Gon{\c c}alves}, {Rodet}, {Russeil},
  {Saraceno}, {Sidher}, {Spencer}, {Swinyard}, {Ward-Thompson}, \&
  {White}}]{Rodon2010}
{Rod{\'o}n}, J.~A., {Zavagno}, A., {Baluteau}, J.-P., {et~al.} 2010, \aap, 518,
  L80+

\bibitem[{{Schnee} {et~al.}(2010){Schnee}, {Enoch}, {Noriega-Crespo}, {Sayers},
  {Terebey}, {Caselli}, {Foster}, {Goodman}, {Kauffmann}, {Padgett}, {Rebull},
  {Sargent}, \& {Shetty}}]{Schnee2010}
{Schnee}, S., {Enoch}, M., {Noriega-Crespo}, A., {et~al.} 2010, \apj, 708, 127

\bibitem[{{Schnee} {et~al.}(2008){Schnee}, {Li}, {Goodman}, \&
  {Sargent}}]{Schnee2008}
{Schnee}, S., {Li}, J., {Goodman}, A.~A., \& {Sargent}, A.~I. 2008, \apj, 684,
  1228

\bibitem[{{Shetty} {et~al.}(2009){Shetty}, {Kauffmann}, {Schnee}, {Goodman}, \&
  {Ercolano}}]{Shetty2009}
{Shetty}, R., {Kauffmann}, J., {Schnee}, S., {Goodman}, A., \& {Ercolano}, B.
  2009, \apj, 696, 2234

\bibitem[{{Stamatellos} {et~al.}(2010){Stamatellos}, {Griffin}, {Kirk},
  {Molinari}, {Sibthorpe}, {Ward-Thompson}, {Whitworth}, \&
  {Wilcock}}]{Stamatellos2010}
{Stamatellos}, D., {Griffin}, M.~J., {Kirk}, J.~M., {et~al.} 2010, \mnras, 409,
  12

\bibitem[{{Stamatellos} \& {Whitworth}(2003)}]{Stamatellos2003}
{Stamatellos}, D. \& {Whitworth}, A.~P. 2003, \aap, 407, 941

\bibitem[{{Stamatellos} {et~al.}(2004){Stamatellos}, {Whitworth}, {Andr{\'e}},
  \& {Ward-Thompson}}]{Stamatellos2004}
{Stamatellos}, D., {Whitworth}, A.~P., {Andr{\'e}}, P., \& {Ward-Thompson}, D.
  2004, \aap, 420, 1009

\bibitem[{{Steinacker} {et~al.}(2010){Steinacker}, {Pagani}, {Bacmann}, \&
  {Guieu}}]{Steinacker2010}
{Steinacker}, J., {Pagani}, L., {Bacmann}, A., \& {Guieu}, S. 2010, \aap, 511,
  A9+

\bibitem[{{Stepnik} {et~al.}(2003){Stepnik}, {Abergel}, {Bernard}, {Boulanger},
  {Cambr{\'e}sy}, {Giard}, {Jones}, {Lagache}, {Lamarre}, {Meny}, {Pajot}, {Le
  Peintre}, {Ristorcelli}, {Serra}, \& {Torre}}]{Stepnik2003}
{Stepnik}, B., {Abergel}, A., {Bernard}, J., {et~al.} 2003, \aap, 398, 551

\bibitem[{Stognienko {et~al.}(1995)Stognienko, Henning, \&
  Ossenkopf}]{Stognienko1995}
Stognienko, R., Henning, T., \& Ossenkopf, V. 1995, \aap, 296, 797

\bibitem[{{Veneziani} {et~al.}(2010){Veneziani}, {Ade}, {Bock}, {Boscaleri},
  {Crill}, {de Bernardis}, {De Gasperis}, {de Oliveira-Costa}, {De Troia}, {Di
  Stefano}, {Ganga}, {Jones}, {Kisner}, {Lange}, {MacTavish}, {Masi},
  {Mauskopf}, {Montroy}, {Natoli}, {Netterfield}, {Pascale}, {Piacentini},
  {Pietrobon}, {Polenta}, {Ricciardi}, {Romeo}, \& {Ruhl}}]{Veneziani2010}
{Veneziani}, M., {Ade}, P.~A.~R., {Bock}, J.~J., {et~al.} 2010, \apj, 713, 959

\bibitem[{{Voshchinnikov} {et~al.}(2006){Voshchinnikov}, {Il'in}, {Henning}, \&
  {Dubkova}}]{Voshchinnikov2006}
{Voshchinnikov}, N.~V., {Il'in}, V.~B., {Henning}, T., \& {Dubkova}, D.~N.
  2006, \aap, 445, 167

\bibitem[{{Weingartner} \& {Draine}(2001)}]{Weingartner2001}
{Weingartner}, J.~C. \& {Draine}, B.~T. 2001, \apj, 548, 296

\bibitem[{{Zucconi} {et~al.}(2001){Zucconi}, {Walmsley}, \&
  {Galli}}]{Zucconi2001}
{Zucconi}, A., {Walmsley}, C.~M., \& {Galli}, D. 2001, \aap, 376, 650

\end{thebibliography}
 
\appendix
 
\section{Details regarding the spectral variations in the emissivity spectral index} 
\label{appendix_spectral_variations}

\begin{figure}[!t]
\centerline{
\includegraphics[width=0.42\textwidth]{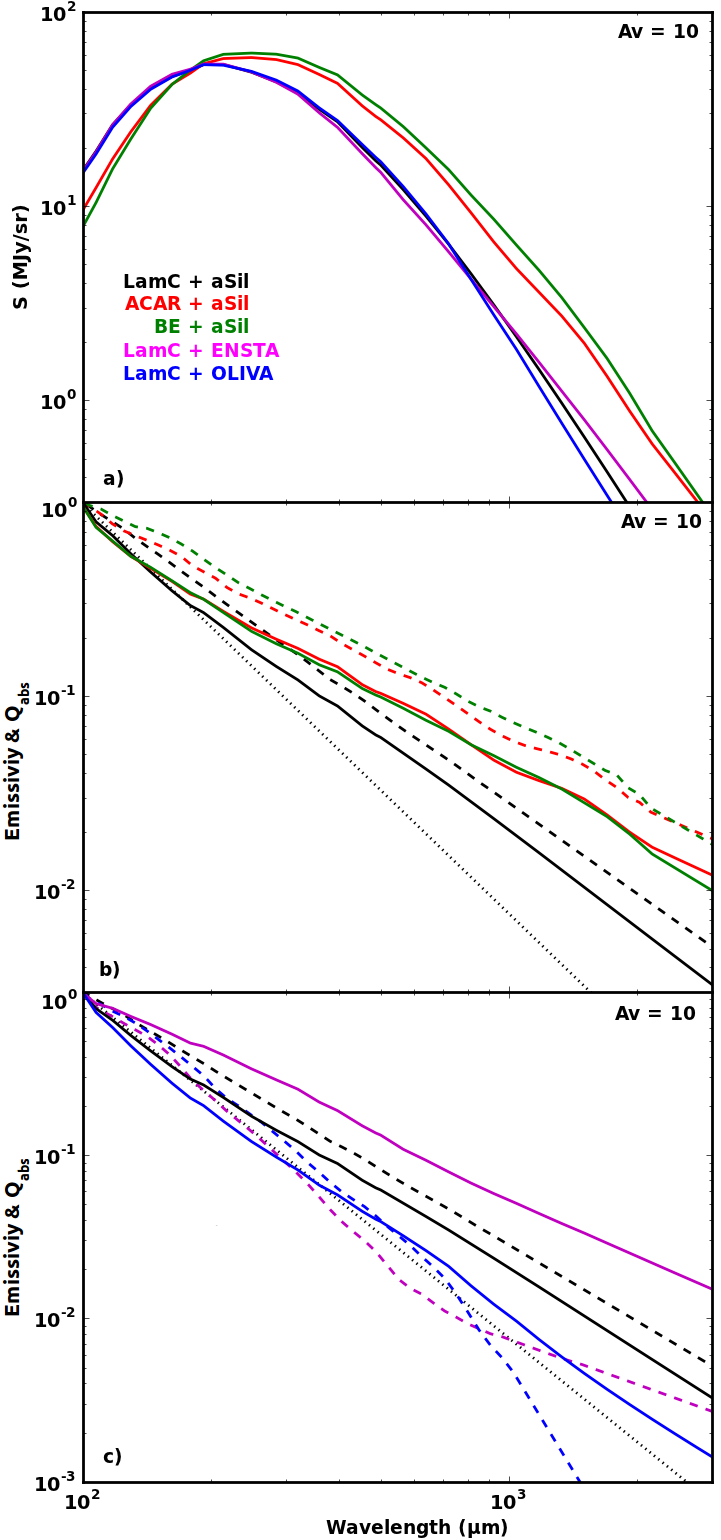}}
\caption{a) SEDs for the central pixel of the cloud with central $A_V = 10$ for a mixture of LamC and aSil (black), LamC and ENSTA (magenta), LamC and OLIVA (blue), ACAR and aSil (red), BE and aSil (green).
b) Emissivity normalised at 100 $\mu$m for the cloud with central $A_V = 10$, for a mixture of LamC and aSil (black solid line), of ACAR and aSil (red solid line), and of BE and aSil (green solid line). Absorption efficiencies normalised at 100 $\mu$m: aSil (black dotted line), LamC (black dashed line), ACAR (red dashed line), and BE (green dashed line).
c) Emissivity normalised at 100 $\mu$m for the cloud with central $A_V = 10$, for a mixture of LamC and aSil (black solid line), of LamC and ENSTA (magenta solid line), and of LamC and OLIVA (blue solid line). Absorption efficiencies normalised at 100 $\mu$m: aSil (black dotted line), LamC (black dashed line), ENSTA (magenta dashed line), and OLIVA (blue dashed line).}
\label{figures_appendix_spectral_variations} 
\end{figure}

\begin{figure}[!th]
\centerline{
\includegraphics[width=0.42\textwidth]{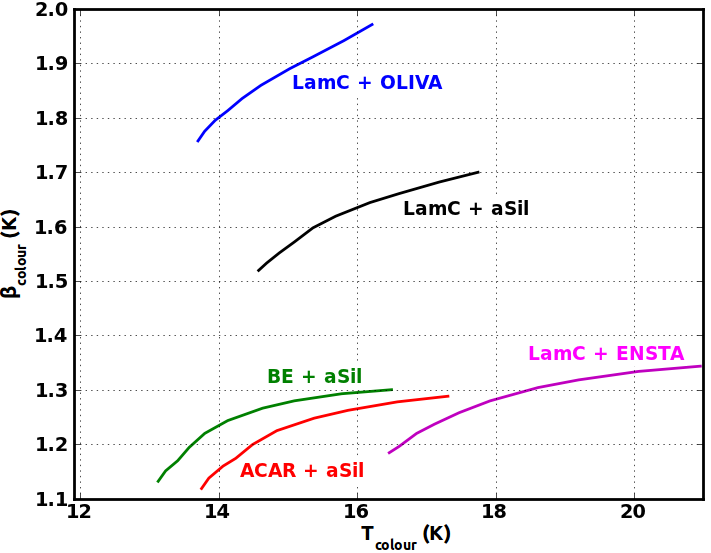}}
\caption{Emissivity spectral index as a function of colour temperature for fits of the dust SEDs between 100 and $3\,000\;\mu$m for the central pixels of the modelled clouds. Linestyles are the same as in Fig. \ref{appendix_spectral_variations}a).}
\label{appendix_beta_T} 
\end{figure}

\normalsize{We} use the ten clouds described in Tab. 3 (diffuse medium case), illuminated by the ISRF, and consider five different mixtures of grains: LamC and aSil (see Tab. 2), ACAR and aSil, BE and aSil, LamC and ENSTA, and LamC and OLIVA (see Section \ref{spectral_variations} for a description of their optical properties). The corresponding SEDs and $(\beta_{colour}-T_{colour})$-parameters, obtained with fits between 100 and $3\,000 \; \mu$m, are shown in Figs. \ref{figures_appendix_spectral_variations}a) and \ref{appendix_beta_T}.

All the mixtures containing LamC grains have similar colour temperatures, $T_{colour}$, regardless of the silicate population (aSil, ENSTA, or OLIVA). If the LamC grains are replaced by ACAR or BE grains, the colour temperature drops. This decrease is due to their higher emissivity in the far-IR and shows that the colour temperature is dominated by the temperature of the carbonaceous populations, which are always the warmest.

The populations of ENSTA and OLIVA grains have very different optical properties in the submillimetre (Fig. \ref{beta_lambda_1}b). This reappears in the fitted values of the emissivity spectral index, $\beta_{colour}$, which are much higher for a mixture containing OLIVA grains ($1.75 \leqslant \beta_{colour} \leqslant 2$) than for a mixture containing ENSTA grains ($1.15 \leqslant \beta_{colour} \leqslant 1.35$, see Fig. \ref{appendix_beta_T}).

Finally, Figs. \ref{figures_appendix_spectral_variations}b) and c) show the spectral variations in the emissivities of the different mixtures of grains, which are measured using Eq. \ref{equation_emissivity}. For all the mixtures containing aSil grains, for $\lambda \geqslant 500 \; \mu$m, the spectral variations in the emissivity perfectly reflect the spectral variations in the absorption efficiencies, $Q_{abs}$, of the carbonaceous grains. For the mixture containing LamC and OLIVA grains, the carbonaceous grains still dominate the spectral variations in the emissivity for $\lambda \geqslant 1\,000 \; \mu$m. However, if we replace the OLIVA by ENSTA grains, this is no longer true and the population of silicates dominates the spectral variations for $\lambda \geqslant 1\,000 \; \mu$m. This illustrates the strong dependence of the emissivity spectral variations on the dust populations considered. However, the relatively simple behaviours described here are all mixed up if we consider that the optical properties of both the silicates and the carbons undergo spectral variations in the far-IR and in the submillimetre (see Section \ref{spectral_variations} for a detailed description). As a result, it might be quite difficult to track back the nature of the grains from the spectral variations in the dust SED.

\section{The case of Bonnor-Ebert spheres}
\label{appendix_BE}

To test the influence of the shape of the clouds on our results, we study the case of externally heated Bonnor-Ebert spheres, which are isothermal spheres at hydrostatic equilibrium \citep{Bonnor1956, Ebert1955}. We consider almost critically stable spheres with $\zeta = 6.5$, or similarly $\rho_C/\rho(R) = 14.1$, where $\zeta$ is the dimensionless length equal to $(4\pi G\rho_C/c_S^2)^{1/2}r$ and $c_S$ is the isothermal sound speed, $G$ the gravitational constant, and $r$ the radius. We assume a constant kinetic gas temperature of $T_{gas} = 12$ K. \citet{Fischera2011} showed that the choice of $\zeta$ and $T_{gas}$ has a minor influence on the dust temperature, which depends mainly on the column density. The modelled clouds are divided into 50 concentric cells (see Section \ref{description_CRT}) and the emission maps produced have $85 \times 85$ pixels with a single pixel size equal to 0.6\% of the radius of the cloud. We vary the masses of the spheres between 0.2 and 100 $M_\odot$ to get central hydrogen column densities ranging from $2.5 \times 10^{20}$ to $5 \times 10^{22}$ H/cm$^{-2}$, or central $A_V = 0.1$ to 17.7 in the case of DHGL grain populations. The parameters describing the spheres are shown in Tab. B.1.

\begin{figure}[!th]
\centerline{
\includegraphics[width=0.42\textwidth]{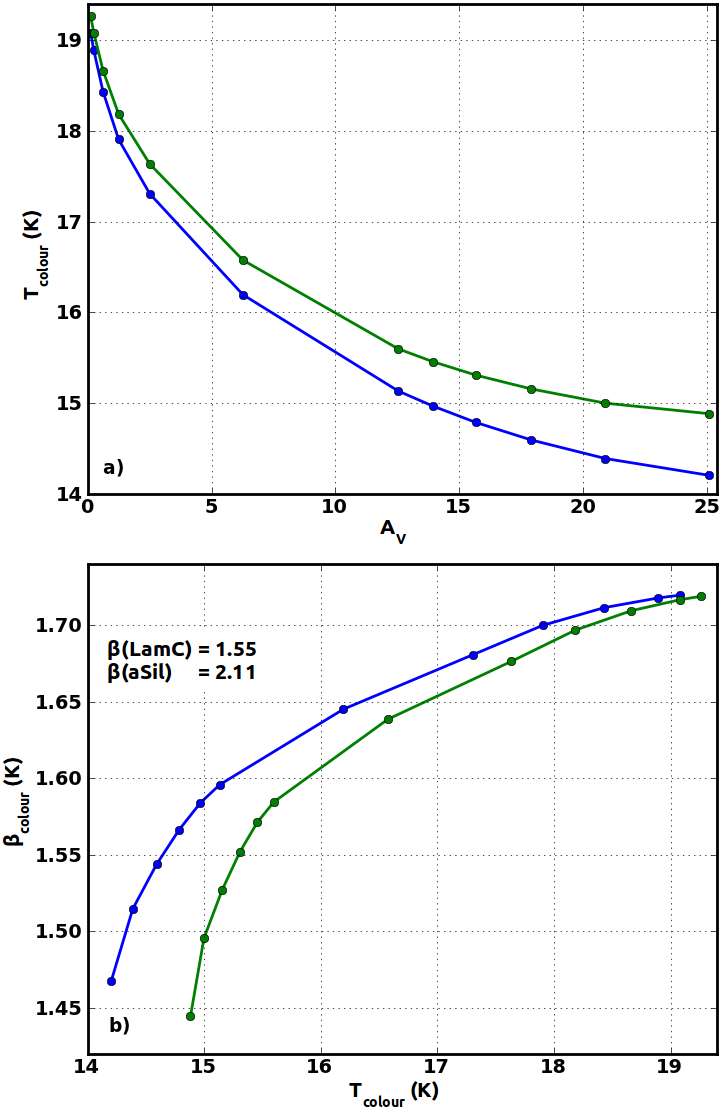}}
\caption{a) Same as Fig. \ref{diffuse_test_case}c in the case of Bonnor-Ebert spheres. b) Same as Fig. \ref{diffuse_test_case}d in the case of Bonnor-Ebert spheres.}
\label{DHGL_BE} 
\end{figure}

\begin{figure}[!th]
\centerline{
\includegraphics[width=0.42\textwidth]{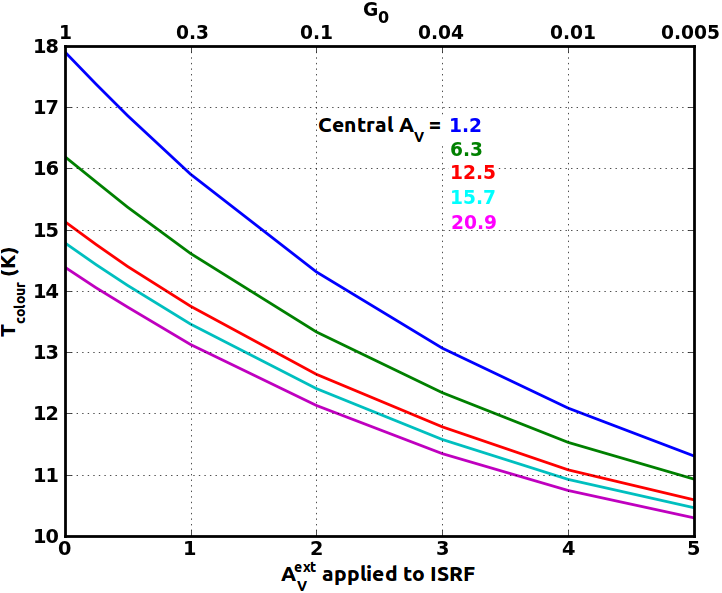}}
\caption{Same as Fig. \ref{T_Go} in the case of Bonnor-Ebert spheres. The blue line shows a sphere with $A_V = 1.2$ at the centre, green shows 6.3, red 12.5, and magenta 20.9.}
\label{extinction_BE} 
\end{figure}

\begin{figure}[!th]
\centerline{
\includegraphics[width=0.42\textwidth]{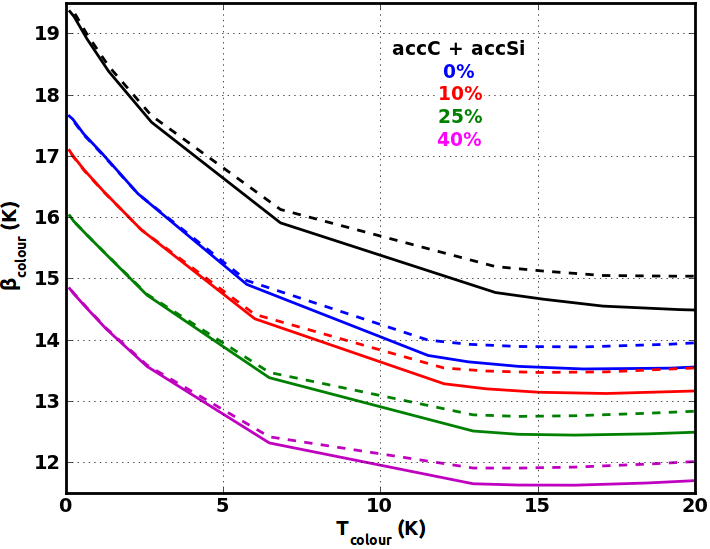}}
\caption{Same as Fig. \ref{grain_growth}a in the case of Bonnor-Ebert spheres.}
\label{growth_BE} 
\end{figure}

For comparison with the cylindrical clouds presented in the main body of the paper, we measure the $\beta_{colour}$ and $T_{colour}$ parameters for Bonnor-Ebert spheres with DHGL dust populations illuminated by the ISRF (Fig. \ref{DHGL_BE}), or by the ISRF extinguished by an external radiation field $A_V^{ext}$ up to 5 (Fig. \ref{extinction_BE}). Figure \ref{growth_BE} shows the case where Bonnor-Ebert spheres, illuminated by the ISRF, contain accC and accSi populations or aggregates with various porosity fractions. The values of the colour temperatures and the emissivity spectral indices vary when compared to the case of cylindrical clouds, although in all cases, the trends are similar. This means that the results described in the main body of the text in the case of cylinders hold for externally heated Bonnor-Ebert spheres.

\begin{figure}[!th]
\centerline{
\includegraphics[width=0.35\textwidth]{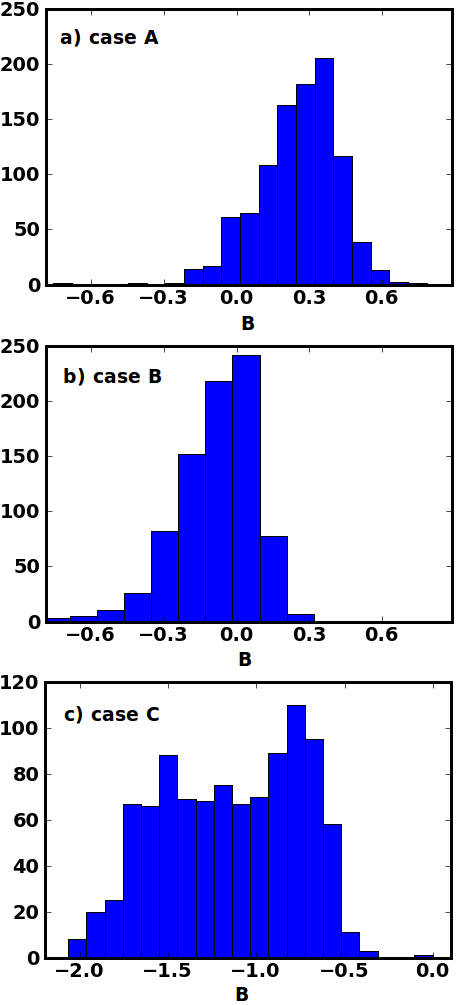}}
\caption{Same as Fig. \ref{beta_T_noise_herschel}c in the case of Bonnor-Ebert spheres, for a signal-to-noise ration at 250 $\mu$m greater than a) 8.3 (case A), b) 4.2 (case B), and c) 1.7 (case C).}
\label{noise_BE} 
\end{figure}

We now study the effects of noise on the $\beta_{colour}-T_{colour}$ relations measured in Bonnor-Ebert spheres for observations in Herschel channels (PACS and SPIRE). The procedure is exactly the same as in Section \ref{section_noise} but we examine three cases depending on the 250 $\mu$m peak surface brightness after convolution with the beam (the column density range covered by the modelled Bonnor-Ebert spheres is indeed larger than for the cylinders). For case A, we exclude the clouds with S/N at 250 $\mu$m lower than 8.3 ($\Leftrightarrow S_{250 \; \mu{\rm m}} < 10$ MJy/sr, excludes all clouds with $M \geqslant 20 \; M_\odot$), for case B lower than 4.2 ($\Leftrightarrow S_{250 \; \mu{\rm m}} < 5$ MJy/sr, $M \geqslant 50 \; M_\odot$), and for case C lower than 1.7 ($\Leftrightarrow S_{250 \; \mu{\rm m}} < 2$ MJy/sr, $M \geqslant 100 \; M_\odot$). The results are presented in Fig. \ref{noise_BE}. In case A, the correlation between $\beta_{colour}$ and $T_{colour}$ remains positive. In case B, a slight anti-correlation appears ($B \geqslant -0.6$), but cannot account alone for the steep anti-correlation observed towards interstellar clouds ($B \leqslant -1$). In case C, where we include the nine modelled clouds with the lowest S/N's, an anti-correlation appears and is steeper than $B = -1$ for more than half of the simulated cloud samples. The few faint clouds with a low S/N are responsible for the strong anti-correlation measured.

\begin{table*}
\label{tableau_BE}
\centering
\caption{Parameters describing the gas distribution of the Bonnor-Ebert spheres. $M$ is the the cloud mass in units of solar masses, $N_H$ is the hydrogen column density at the centre of the cloud, $\rho_C$ is the central hydrogen density, $R$ is the external radius of the sphere, and $A_V$ is the visual extinction at the centre of the cloud, which depends on the grain populations considered (labels are the same as in Tab. 2).}
\begin{tabular}{cccccccccc}
\hline
\hline
$M$         & $N_H$               & $\rho_C$   & $R$  & $A_V$ & $A_V$    & $A_V$ & $A_V$ & $A_V$ & $A_V$ \\
($M_\odot$) & (H/cm$^2$)          & (H/cm$^3$) & (pc) & DHGL  & accretion & 0\%   & 10\%  & 25\%  & 40\% \\
\hline
0.5         & $5\times 10^{22}$   & 780072     & 0.025 & 25.1 & 27.3 & 23.0 & 24.0 & 25.9 & 27.9 \\
0.6         & $4\times 10^{22}$   & 541717     & 0.03  & 20.9 & 22.8 & 19.2 & 20.0 & 21.6 & 23.3 \\
0.7         & $3.5\times 10^{22}$ & 397996     & 0.035 & 17.9 & 19.5 & 16.5 & 17.2 & 18.5 & 19.9 \\
0.8         & $3\times 10^{22}$   & 304716     & 0.04  & 15.7 & 17.1 & 14.4 & 15.0 & 16.2 & 17.5 \\
0.9         & $2.7\times 10^{22}$ & 240762     & 0.045 & 13.9 & 15.2 & 12.8 & 13.4 & 14.4 & 15.5 \\
1           & $2.5\times 10^{22}$ & 195018     & 0.05  & 12.5 & 13.7 & 11.5 & 12.0 & 12.9 & 14.0 \\
2           & $1.2\times 10^{22}$ & 48755      & 0.1   & 6.3  & 6.8  & 5.8  & 6.0  & 6.5  & 7.0 \\
5           & $5\times 10^{21}$   & 7801       & 0.25  & 2.5  & 2.7  & 2.3  & 2.4  & 2.6  & 2.8 \\
10          & $2.5\times 10^{21}$ & 1950       & 0.5   & 1.2  & 1.4  & 1.2  & 1.2  & 1.3  & 1.4 \\
20          & $1.2\times 10^{21}$ & 488        & 1.0   & 0.6  & 0.7  & 0.6  & 0.6  & 0.7  & 0.7 \\
50          & $5\times 10^{20}$   & 78         & 2.5   & 0.3  & 0.3  & 0.2  & 0.2  & 0.3  & 0.3 \\
100         & $2.5\times 10^{20}$ & 20         & 5.0   & 0.1  & 0.1  & 0.1  & 0.1  & 0.1  & 0.1 \\
\hline
\end{tabular}
\end{table*}

\end{document}